\input harvmac
\def\figflag{I}
\noblackbox

\font\cmss=cmss10 \font\cmsss=cmss10 at 7pt
 \def\inbar{\,\vrule height1.5ex width.4pt depth0pt}
\def\IZ{\relax\ifmmode\mathchoice
{\hbox{\cmss Z\kern-.4em Z}}{\hbox{\cmss Z\kern-.4em Z}} {\lower.9pt\hbox{\cmsss Z\kern-.4em Z}}
{\lower1.2pt\hbox{\cmsss Z\kern-.4em Z}}\else{\cmss Z\kern-.4em Z}\fi}
\def\IB{\relax{\rm I\kern-.18em B}}
\def\IC{{\relax\hbox{$\inbar\kern-.3em{\rm C}$}}}
\def\ID{\relax{\rm I\kern-.18em D}}
\def\IE{\relax{\rm I\kern-.18em E}}
\def\IF{\relax{\rm I\kern-.18em F}}
\def\IG{\relax\hbox{$\inbar\kern-.3em{\rm G}$}}
\def\IGa{\relax\hbox{${\rm I}\kern-.18em\Gamma$}}
\def\IH{\relax{\rm I\kern-.18em H}}
\def\II{\relax{\rm I\kern-.18em I}}
\def\IK{\relax{\rm I\kern-.18em K}}
\def\IR{\relax{\rm I\kern-.18em R}}

\def\tu{\tilde u}
\def\tx{\tilde x}

\def\tfig#1{Figure~\the\figno\xdef#1{Figure~\the\figno}\global\advance\figno by1}
\def\figI{I}
%
\newdimen\tempszb \newdimen\tempszc \newdimen\tempszd \newdimen\tempsze
\ifx\figflag\figI
\input epsf
%
%
%
%
%
\def\ifigure#1#2#3#4{
\midinsert \vbox to #4truein{\ifx\figflag\figI \vfil\centerline{\epsfysize=#4truein\epsfbox{#3}}\fi}
\baselineskip=12pt \narrower\narrower\noindent{\bf #1:} #2
\endinsert
}


\lref\Allan{
  A.~Adams, M.~Ernebjerg and J.~M.~Lapan,
  ``Linear models for flux vacua,''
  arXiv:hep-th/0611084.
}

\lref\SharpeWU{
  E.~R.~Sharpe,
  ``Recent developments in discrete torsion,''
  Phys.\ Lett.\  B {\bf 498}, 104 (2001)
  [arXiv:hep-th/0008191].
}

\lref\FreyPol{
  A.~R.~Frey and J.~Polchinski,
  ``N = 3 warped compactifications,''
  Phys.\ Rev.\  D {\bf 65}, 126009 (2002)
  [arXiv:hep-th/0201029].
}

\lref\WestphalXD{
  A.~Westphal,
  ``Lifetime of Stringy de Sitter Vacua,''
  arXiv:0705.1557 [hep-th].
}

\lref\GukovYA{
  S.~Gukov, C.~Vafa and E.~Witten,
  ``CFT's from Calabi-Yau four-folds,''
  Nucl.\ Phys.\  B {\bf 584}, 69 (2000)
  [Erratum-ibid.\  B {\bf 608}, 477 (2001)]
  [arXiv:hep-th/9906070].
}

\lref\KKmon{
  D.~J.~Gross and M.~J.~Perry,
  ``Magnetic Monopoles In Kaluza-Klein Theories,''
  Nucl.\ Phys.\  B {\bf 226}, 29 (1983);
  R.~d.~Sorkin,
  ``Kaluza-Klein Monopole,''
  Phys.\ Rev.\ Lett.\  {\bf 51}, 87 (1983).
}

\lref\SaueressigES{
  F.~Saueressig, U.~Theis and S.~Vandoren,
  ``On de Sitter vacua in type IIA orientifold compactifications,''
  Phys.\ Lett.\  B {\bf 633}, 125 (2006)
  [arXiv:hep-th/0506181].
}

\lref\DT{ C.~Vafa,
  ``Modular Invariance And Discrete Torsion On Orbifolds,''
  Nucl.\ Phys.\  B {\bf 273}, 592 (1986).
}

\lref\HouseYC{
  T.~House and E.~Palti,
  ``Effective action of (massive) IIA on manifolds with SU(3) structure,''
  Phys.\ Rev.\  D {\bf 72}, 026004 (2005)
  [arXiv:hep-th/0505177].
}

\lref\CamaraDC{
  P.~G.~Camara, A.~Font and L.~E.~Ibanez,
  ``Fluxes, moduli fixing and MSSM-like vacua in a simple IIA orientifold,''
  JHEP {\bf 0509}, 013 (2005)
  [arXiv:hep-th/0506066].
}

\lref\GranaKF{ D.~Lust and D.~Tsimpis,
  ``Supersymmetric AdS(4) compactifications of IIA supergravity,''
  JHEP {\bf 0502}, 027 (2005)
  [arXiv:hep-th/0412250];
  M.~Grana, R.~Minasian, M.~Petrini and A.~Tomasiello,
  ``A scan for new N=1 vacua on twisted tori,''
  JHEP {\bf 0705}, 031 (2007)
  [arXiv:hep-th/0609124].
}

\lref\SheltonFD{
  J.~Shelton, W.~Taylor and B.~Wecht,
  ``Generalized flux vacua,''
  JHEP {\bf 0702}, 095 (2007)
  [arXiv:hep-th/0607015].
}

\lref\Wraseetal{
  D.~Robbins and T.~Wrase,
  ``D-Terms from Generalized NS-NS Fluxes in Type II,''
  arXiv:0709.2186 [hep-th].
  M.~Ihl, D.~Robbins and T.~Wrase,
  ``Toroidal Orientifolds in IIA with General NS-NS Fluxes,''
  JHEP {\bf 0708}, 043 (2007)
  [arXiv:0705.3410 [hep-th]].
  M.~Ihl and T.~Wrase,
  ``Towards a realistic type IIA T**6/Z(4) orientifold model with background
  fluxes. I: Moduli stabilization,''
  JHEP {\bf 0607}, 027 (2006)
  [arXiv:hep-th/0604087].
}

\lref\albion{A.~Lawrence, T.~Sander, M.~B.~Schulz and B.~Wecht,
  ``Torsion and Supersymmetry Breaking,''
  arXiv:0711.4787 [hep-th].
}

\lref\VafaWitten{ C.~Vafa and E.~Witten,
  ``On orbifolds with discrete torsion,''
  J.\ Geom.\ Phys.\  {\bf 15}, 189 (1995)
  [arXiv:hep-th/9409188].
}

\lref\KKNS{ R.~Gregory, J.~A.~Harvey and G.~W.~Moore,
  ``Unwinding strings and T-duality of Kaluza-Klein and H-monopoles,''
  Adv.\ Theor.\ Math.\ Phys.\  {\bf 1}, 283 (1997)
  [arXiv:hep-th/9708086];
  D. Tong,
``NS5-branes, T-duality and worldsheet instantons,''
  JHEP {\bf 0207}, 013 (2002)
  [arXiv:hep-th/0204186];
H.~Ooguri and C.~Vafa,
  ``Two-Dimensional Black Hole and Singularities of CY Manifolds,''
  Nucl.\ Phys.\ B {\bf 463}, 55 (1996)
  [arXiv:hep-th/9511164].
}

\lref\VZ{ G.~Villadoro and F.~Zwirner,
  ``On general flux backgrounds with localized sources,''
  arXiv:0710.2551 [hep-th];
  G.~Villadoro and F.~Zwirner,
  ``Beyond Twisted Tori,''
  Phys.\ Lett.\  B {\bf 652}, 118 (2007)
  [arXiv:0706.3049 [hep-th]].
}

\lref\FreedWitten{ R.~Minasian and G.~W.~Moore,
  ``K-theory and Ramond-Ramond charge,''
  JHEP {\bf 9711}, 002 (1997)
  [arXiv:hep-th/9710230];
  E.~Witten,
  ``D-branes and K-theory,''
  JHEP {\bf 9812}, 019 (1998)
  [arXiv:hep-th/9810188].
D.~S.~Freed and E.~Witten,
  ``Anomalies in string theory with D-branes,''
  arXiv:hep-th/9907189;
  R.~Minasian and G.~W.~Moore,
  ``K-theory and Ramond-Ramond charge,''
  JHEP {\bf 9711}, 002 (1997)
  [arXiv:hep-th/9710230].
}

\lref\Ktheory{See e.g. G.~W.~Moore and E.~Witten,
  ``Self-duality, Ramond-Ramond fields, and K-theory,''
  JHEP {\bf 0005}, 032 (2000)
  [arXiv:hep-th/9912279];
  A.~M.~Uranga,
  ``D-brane probes, RR tadpole cancellation and K-theory charge,''
  Nucl.\ Phys.\  B {\bf 598}, 225 (2001)
  [arXiv:hep-th/0011048].
}

\lref\thurston{W.P. Thurston, ``Three-Dimensional Geometry and Topology", Princeton Univ. Press (1997).}

\lref\mit{M.~P.~Hertzberg, S.~Kachru, W.~Taylor and M.~Tegmark,
  ``Inflationary Constraints on Type IIA String Theory,''
  arXiv:0711.2512 [hep-th];
 M.~P.~Hertzberg, M.~Tegmark, S.~Kachru, J.~Shelton and O.~Ozcan,
  ``Searching for Inflation in Simple String Theory Models: An Astrophysical
  Perspective,''
  arXiv:0709.0002 [astro-ph].
}

\lref\junctions{O. Aharony, Y. Antebi and M. Berkooz, to appear; E.~Silverstein,
  ``AdS and dS entropy from string junctions or the function of junction
  conjunctions,''
  arXiv:hep-th/0308175.
}

\lref\BanksHG{
  T.~Banks and K.~van den Broek,
  ``Massive IIA flux compactifications and U-dualities,''
  JHEP {\bf 0703}, 068 (2007)
  [arXiv:hep-th/0611185].
}
\lref\VilladoroCU{
  G.~Villadoro and F.~Zwirner,
  ``N = 1 effective potential from dual type-IIA D6/O6 orientifolds with
  general fluxes,''
  JHEP {\bf 0506}, 047 (2005)
  [arXiv:hep-th/0503169].
}

\lref\dualgen{
  A.~Strominger,
 ``Inflation and the dS/CFT correspondence,''
  JHEP {\bf 0111}, 049 (2001)
  [arXiv:hep-th/0110087].

  T.~Banks and W.~Fischler,
 ``An holographic cosmology,''
  arXiv:hep-th/0111142.

  M.~Alishahiha, A.~Karch, E.~Silverstein and D.~Tong,
 ``The dS/dS correspondence,''
  AIP Conf.\ Proc.\  {\bf 743}, 393 (2005)
  [arXiv:hep-th/0407125].

  M.~Alishahiha, A.~Karch and E.~Silverstein,
 ``Hologravity,''
  JHEP {\bf 0506}, 028 (2005)
  [arXiv:hep-th/0504056].

  B.~Freivogel, V.~E.~Hubeny, A.~Maloney, R.~Myers, M.~Rangamani and S.~Shenker,
 ``Inflation in AdS/CFT,''
  JHEP {\bf 0603}, 007 (2006)
  [arXiv:hep-th/0510046].

  B.~Freivogel, Y.~Sekino, L.~Susskind and C.~P.~Yeh,
 ``A holographic framework for eternal inflation,''
  Phys.\ Rev.\  D {\bf 74}, 086003 (2006)
  [arXiv:hep-th/0606204].

  N.~Arkani-Hamed, S.~Dubovsky, A.~Nicolis, E.~Trincherini and G.~Villadoro,
  ``A Measure of de Sitter Entropy and Eternal Inflation,''
  JHEP {\bf 0705}, 055 (2007)
  [arXiv:0704.1814 [hep-th]].
}

\lref\Inflationreviews{
  S.~H.~Henry Tye,
  ``Brane inflation: String theory viewed from the cosmos,''
  arXiv:hep-th/0610221;

J.~M.~Cline,
  ``String cosmology,''
  arXiv:hep-th/0612129;

R.~Kallosh,
  ``On Inflation in String Theory,''
  arXiv:hep-th/0702059;

C.~P.~Burgess,
 ``Lectures on Cosmic Inflation and its Potential Stringy Realizations,''
  arXiv:0708.2865 [hep-th];
L.~McAllister and E.~Silverstein,
  ``String Cosmology: A Review,''
  arXiv:0710.2951 [hep-th].
}

\lref\axionpot{
  Y.~Hosotani,
  ``Dynamical Mass Generation By Compact Extra Dimensions,''
  Phys.\ Lett.\  B {\bf 126}, 309 (1983);
  H.~Hatanaka, T.~Inami and C.~S.~Lim,
  ``The gauge hierarchy problem and higher dimensional gauge theories,''
  Mod.\ Phys.\ Lett.\  A {\bf 13}, 2601 (1998)
  [arXiv:hep-th/9805067];
  I.~Antoniadis, K.~Benakli and M.~Quiros,
  ``Finite Higgs mass without supersymmetry,''
  New J.\ Phys.\  {\bf 3}, 20 (2001)
  [arXiv:hep-th/0108005].
  G.~von Gersdorff, N.~Irges and M.~Quiros,
  Nucl.\ Phys.\  B {\bf 635}, 127 (2002)
  [arXiv:hep-th/0204223];
  H.~C.~Cheng, K.~T.~Matchev and M.~Schmaltz,
  ``Radiative corrections to Kaluza-Klein masses,''
  Phys.\ Rev.\  D {\bf 66}, 036005 (2002)
  [arXiv:hep-ph/0204342].
  N.~Arkani-Hamed, H.~C.~Cheng, P.~Creminelli and L.~Randall,
  ``Pseudonatural inflation,''
  JCAP {\bf 0307}, 003 (2003)
  [arXiv:hep-th/0302034].
}

\lref\alex{A. Maloney, unpublished; A. Lawrence, A. Maloney, E. Silverstein, and D. Starr, works in progress.}

\lref\IBM{
  R.~Blumenhagen, M.~Cvetic, P.~Langacker and G.~Shiu,
  ``Toward realistic intersecting D-brane models,''
  Ann.\ Rev.\ Nucl.\ Part.\ Sci.\  {\bf 55}, 71 (2005)
  [arXiv:hep-th/0502005].
}

\lref\gromov{M. Gromov, ``Groups of Polynomial growth and Expanding Maps", Publications mathematiques I.H.É.S.,
53, 1981}

\lref\fractCS{S.~Gukov, S.~Kachru, X.~Liu and L.~McAllister,
  ``Heterotic moduli stabilization with fractional Chern-Simons invariants,''
  Phys.\ Rev.\  D {\bf 69}, 086008 (2004)
  [arXiv:hep-th/0310159].
}

\lref\dualsing{ T.~Banks, W.~Fischler and L.~Motl,
  ``Dualities versus singularities,''
  JHEP {\bf 9901}, 019 (1999)
  [arXiv:hep-th/9811194].
}

\lref\robbins{ D.~Robbins and T.~Wrase,
  ``D-Terms from Generalized NS-NS Fluxes in Type II,''
  arXiv:0709.2186 [hep-th].
}

\lref\WLfiber{ T.~Ishii, G.~Ishiki, S.~Shimasaki and A.~Tsuchiya,
  ``T-duality, fiber bundles and matrices,''
  JHEP {\bf 0705}, 014 (2007)
  [arXiv:hep-th/0703021].
}

\lref\GSW{
  M.~B.~Green, J.~H.~Schwarz and E.~Witten,
{\it  Cambridge, Uk: Univ. Pr. ( 1987) 596 P. ( Cambridge Monographs On Mathematical Physics)} }

\lref\reviews{ A.~R.~Frey,
  ``Warped strings: Self-dual flux and contemporary compactifications,''
  arXiv:hep-th/0308156;
E.~Silverstein,
  ``TASI / PiTP / ISS lectures on moduli and microphysics,''
  arXiv:hep-th/0405068;
M.~Grana,
  ``Flux compactifications in string theory: A comprehensive review,''
  Phys.\ Rept.\  {\bf 423}, 91 (2006)
  [arXiv:hep-th/0509003];
  J.~Polchinski,
  ``The cosmological constant and the string landscape,''
  arXiv:hep-th/0603249;
M.~R.~Douglas and S.~Kachru,
  ``Flux compactification,''
  Rev.\ Mod.\ Phys.\  {\bf 79}, 733 (2007)
  [arXiv:hep-th/0610102];
R.~Bousso,
  ``TASI Lectures on the Cosmological Constant,''
  arXiv:0708.4231 [hep-th].
}

\lref\MaloneyRR{
  A.~Maloney, E.~Silverstein and A.~Strominger,
    ``De Sitter space in noncritical string theory,''
    Cambridge 2002, The future of theoretical physics and cosmology* 570-591
[arXiv:hep-th/0205316].
}

\lref\BalasubramanianZX{
  V.~Balasubramanian, P.~Berglund, J.~P.~Conlon and F.~Quevedo,
  ``Systematics of moduli stabilisation in Calabi-Yau flux
  compactifications,''
  JHEP {\bf 0503}, 007 (2005)
  [arXiv:hep-th/0502058].
}

\lref\Dduality{ 
  D.~Green, A.~Lawrence, J.~McGreevy, D.~R.~Morrison and E.~Silverstein,
  ``Dimensional Duality,''
  Phys.\ Rev.\  D {\bf 76}, 066004 (2007)
  [arXiv:0705.0550 [hep-th]].

  J.~McGreevy, E.~Silverstein and D.~Starr,
  ``New dimensions for wound strings: The modular transformation of geometry to
  topology,''
  Phys.\ Rev.\  D {\bf 75}, 044025 (2007)
  [arXiv:hep-th/0612121].

E.~Silverstein,
  ``Dimensional mutation and spacelike singularities,''
  Phys.\ Rev.\  D {\bf 73}, 086004 (2006)
  [arXiv:hep-th/0510044].
}

\lref\SaltmanJH{
  A.~Saltman and E.~Silverstein,
    ``A new handle on de Sitter compactifications,''
      JHEP {\bf 0601}, 139 (2006)
        [arXiv:hep-th/0411271].
      }

\lref\KachruAW{
  S.~Kachru, R.~Kallosh, A.~Linde and S.~P.~Trivedi,
    ``De Sitter vacua in string theory,''
      Phys.\ Rev.\  D {\bf 68}, 046005 (2003)
        [arXiv:hep-th/0301240].
      }

\lref\BoussoXA{
  R.~Bousso and J.~Polchinski,
  ``Quantization of four-form fluxes and dynamical neutralization of the
  cosmological constant,''
  JHEP {\bf 0006}, 006 (2000)
  [arXiv:hep-th/0004134].
}

\lref\SilversteinXN{
  E.~Silverstein,
  ``(A)dS backgrounds from asymmetric orientifolds,''
  arXiv:hep-th/0106209.
}

\lref\GiddingsYU{
  S.~B.~Giddings, S.~Kachru and J.~Polchinski,
  ``Hierarchies from fluxes in string compactifications,''
  Phys.\ Rev.\  D {\bf 66}, 106006 (2002)
  [arXiv:hep-th/0105097].
}

\lref\BeckerPM{ K.~Becker and M.~Becker,
  ``M-Theory on Eight-Manifolds,''
  Nucl.\ Phys.\  B {\bf 477}, 155 (1996)
  [arXiv:hep-th/9605053];
 K.~Becker and M.~Becker,
  ``Supersymmetry breaking, M-theory and fluxes,''
  JHEP {\bf 0107}, 038 (2001)
  [arXiv:hep-th/0107044].
}

\lref\DenefPQ{
  F.~Denef, M.~R.~Douglas and S.~Kachru,
  ``Physics of string flux compactifications,''
  arXiv:hep-th/0701050.
}

\lref\DenefMM{
  F.~Denef, M.~R.~Douglas, B.~Florea, A.~Grassi and S.~Kachru,
  ``Fixing all moduli in a simple F-theory compactification,''
  Adv.\ Theor.\ Math.\ Phys.\  {\bf 9}, 861 (2005)
  [arXiv:hep-th/0503124].
}

\lref\DeWolfeUU{
  O.~DeWolfe, A.~Giryavets, S.~Kachru and W.~Taylor,
  ``Type IIA moduli stabilization,''
  JHEP {\bf 0507}, 066 (2005)
  [arXiv:hep-th/0505160].
}

\lref\KachruSK{
  S.~Kachru, M.~B.~Schulz, P.~K.~Tripathy and S.~P.~Trivedi,
  ``New supersymmetric string compactifications,''
  JHEP {\bf 0303}, 061 (2003)
  [arXiv:hep-th/0211182].
}

\lref\Iprime{
  J.~Polchinski and E.~Witten,
  Nucl.\ Phys.\  B {\bf 460}, 525 (1996)
  [arXiv:hep-th/9510169].
}

\lref\Othernil{Other refs on nilmanifold compactifications...}

\lref\SS{
  J.~Scherk and J.~H.~Schwarz,
  ``How To Get Masses From Extra Dimensions,''
  Nucl.\ Phys.\  B {\bf 153}, 61 (1979);
  N.~Kaloper and R.~C.~Myers,
  ``The O(dd) story of massive supergravity,''
  JHEP {\bf 9905}, 010 (1999)
  [arXiv:hep-th/9901045].
C.~M.~Hull and R.~A.~Reid-Edwards,
  ``Flux compactifications of string theory on twisted tori,''
  arXiv:hep-th/0503114.
}

\lref\GibbonsMU{
  G.~W.~Gibbons and S.~W.~Hawking,
  ``Cosmological Event Horizons, Thermodynamics, And Particle Creation,''
  Phys.\ Rev.\  D {\bf 15}, 2738 (1977).
}

\lref\DineGX{
  M.~Dine, A.~Morisse, A.~Shomer and Z.~Sun,
  ``IIA moduli stabilization with badly broken supersymmetry,''
  arXiv:hep-th/0612189.
}

\lref\DineHE{
  M.~Dine and N.~Seiberg,
  ``Is The Superstring Weakly Coupled?,''
  Phys.\ Lett.\  B {\bf 162}, 299 (1985).
}

\lref\GiryavetsVD{
  A.~Giryavets, S.~Kachru, P.~K.~Tripathy and S.~P.~Trivedi,
  ``Flux compactifications on Calabi-Yau threefolds,''
  JHEP {\bf 0404}, 003 (2004)
  [arXiv:hep-th/0312104].
}

\lref\otherkklt{
 V.~Balasubramanian, P.~Berglund, J.~P.~Conlon and F.~Quevedo,
  ``Systematics of moduli stabilisation in Calabi-Yau flux
  compactifications,''
  JHEP {\bf 0503}, 007 (2005)
  [arXiv:hep-th/0502058];
A.~Saltman and E.~Silverstein,
  ``The scaling of the no-scale potential and de Sitter model building,''
  JHEP {\bf 0411}, 066 (2004)
  [arXiv:hep-th/0402135];
  A.~Misra and P.~Shukla,
  ``Area Codes, Large Volume (Non-)Perturbative alpha'- and Instanton -
  Corrected Non-supersymmetric (A)dS minimum, the Inverse Problem and Fake
  Superpotentials for Multiple-Singular-Loci-Two-Parameter Calabi-Yau's,''
  arXiv:0707.0105 [hep-th].
}

\lref\ConlonDS{
  J.~P.~Conlon and F.~Quevedo,
  ``On the explicit construction and statistics of Calabi-Yau flux vacua,''
  JHEP {\bf 0410}, 039 (2004)
  [arXiv:hep-th/0409215].
}

\lref\DenefDM{
  F.~Denef, M.~R.~Douglas and B.~Florea,
  ``Building a better racetrack,''
  JHEP {\bf 0406}, 034 (2004)
  [arXiv:hep-th/0404257].
}

\lref\KachruSK{
  S.~Kachru, M.~B.~Schulz, P.~K.~Tripathy and S.~P.~Trivedi,
  ``New supersymmetric string compactifications,''
  JHEP {\bf 0303}, 061 (2003)
  [arXiv:hep-th/0211182].
}

\lref\BobkovCY{
  K.~Bobkov,
  ``Volume stabilization via alpha' corrections in type IIB theory with
  fluxes,''
  JHEP {\bf 0505}, 010 (2005)
  [arXiv:hep-th/0412239].
}

\lref\DasguptaSS{
  K.~Dasgupta, G.~Rajesh and S.~Sethi,
  ``M theory, orientifolds and G-flux,''
  JHEP {\bf 9908}, 023 (1999)
  [arXiv:hep-th/9908088].
}

\lref\FengIF{
  J.~L.~Feng, J.~March-Russell, S.~Sethi and F.~Wilczek,
  ``Saltatory relaxation of the cosmological constant,''
  Nucl.\ Phys.\  B {\bf 602}, 307 (2001)
  [arXiv:hep-th/0005276].
}

\lref\GrimmUA{
  T.~W.~Grimm and J.~Louis,
  ``The effective action of type IIA Calabi-Yau orientifolds,''
  Nucl.\ Phys.\  B {\bf 718}, 153 (2005)
  [arXiv:hep-th/0412277].
}

\lref\PolchinskiRR{
  J.~Polchinski,
  ``String theory. Vol. 2: Superstring theory and beyond,''
{\it  Cambridge, UK: Univ. Pr. (1998) 531 p} }


\Title {\vbox{\baselineskip12pt\hbox{SLAC-PUB-13016} \hbox{SU-ITP-07/20} \hbox{}}} {\vbox{
\centerline{Simple de Sitter Solutions}} } \centerline{Eva Silverstein }

%
%
%

\bigskip
\centerline{Department of Physics and SLAC} \centerline{Stanford University} \centerline{Stanford, CA 94305,
USA}
\medskip
\bigskip
\medskip
\noindent We present a framework for de Sitter model building in type IIA string theory, illustrated with
specific examples. We find metastable dS minima of the potential for moduli obtained from a compactification on
a product of two Nil three-manifolds (which have negative scalar curvature) combined with orientifolds, branes,
fractional Chern-Simons forms, and fluxes. As a discrete quantum number is taken large, the curvature, field
strengths, inverse volume, and four dimensional string coupling become parametrically small, and the de Sitter
Hubble scale can be tuned parametrically smaller than the scales of the moduli, KK, and winding mode masses. A
subtle point in the construction is that although the curvature remains consistently weak, the circle fibers of
the nilmanifolds become very small in this limit (though this is avoided in illustrative solutions at modest
values of the parameters).
In the simplest version of the construction, the heaviest moduli masses are parametrically of the same order as
the lightest KK and winding masses. However, we provide a method for separating these marginally overlapping
scales, and more generally the underlying supersymmetry of the model protects against large corrections to the
low-energy moduli potential.

\medskip

\Date{December 2007}


\newsec{Introduction}

Cosmological observations and conceptual questions of quantum gravity motivate string-theoretic models of de
Sitter space and inflation (for reviews, see \refs{\reviews,\Inflationreviews}). Several general classes of
constructions of metastable $dS_4$ have been outlined in different corners of string theory with various scales
of supersymmetry breaking \MaloneyRR\refs{\KachruAW,\otherkklt} \SaltmanJH. These followed earlier work on flux
stabilization such as \refs{\BeckerPM,\GukovYA,\DasguptaSS,\GiddingsYU,\SilversteinXN}\ and the original
realization \refs{\BoussoXA,\FengIF}\ that string theory produces a finely-spaced discretuum of flux
contributions to the moduli potential.

Simple and explicit models of compactification down to ${\bf A}dS_4$ have been found using this general
framework\foot{Some authors, notably T. Banks, have questioned the use of the effective field theories
descending from string theory in backgrounds such as de Sitter or anti de Sitter space which are, globally,
infinitely far away from the flat spacetime or linear dilaton backgrounds in which the effective theories were
originally derived. Moreover, as in \refs{\DeWolfeUU,\GrimmUA,\VilladoroCU}, we will consider massive IIA
supergravity, which does not have an exactly flat spacetime solution and so has not been derived precisely from
a string S-matrix in any background (see \BanksHG\ for an exploration of duality in this context). However the
energy densities in our solution will be small away from defects (whose tensions and charges are well understood
in weakly curved spacetime), and we regard it as a conservative working hypothesis that the effective theory
applies in this regime.} in type IIB string theory \refs{\DenefMM,\DenefDM,\ConlonDS} and IIA
\refs{\DeWolfeUU,\VilladoroCU,\HouseYC,\CamaraDC,\GrimmUA,\GranaKF}. The latter set are particularly appealing,
as they make use of power law effects in the string coupling and inverse radii to stabilize all the moduli in
some examples \DeWolfeUU.  A number of works (e.g. \VZ) have been developing a systematic treatment of the
consistency conditions for the ingredients involved in this class of compactifications (as well as more general
candidate models involving ``nongeometric fluxes" such as \refs{\SheltonFD,\Wraseetal,\albion}). Although some
of the previously outlined de Sitter constructions use only such power law forces \refs{\MaloneyRR,\SaltmanJH},
none attain the explicit simplicity of the known $AdS_4$ models.

In this work, we obtain a reasonably simple and explicit set of metastable ${\bf dS}_4$ minima of the moduli
potential by combining the most basic features of \SaltmanJH\ and
\refs{\DeWolfeUU,\VilladoroCU,\CamaraDC,\GrimmUA,\GranaKF,\VZ}, using classic classical results of \SS. From
\SaltmanJH\ we take the strategy of using negative scalar curvature as a leading positive term in the potential,
but in this case we use a simpler compactification (an orientifold of a product of two Nil three-manifolds). The
curvature energy competes against -- without overwhelming -- the contributions from orientifolds, branes, and RR
fluxes in the subsequent orders in the expansion in the string coupling $g_s$. Nil manifolds are twisted
versions of tori, allowing us to fairly straightforwardly generalize the mechanism employed in \DeWolfeUU\ in
the toroidal orientifold case.

In addition to providing positive potential energy, the geometry -- and corresponding topology -- of our
compactification manifold automatically plays two other very useful roles.  First, in contrast to the zero
curvature case, the curvature yields positive mass squared for some angular metric deformations, an effect which
can be understood from the reduced isometry group of the compactification (which corresponds to a reduced number
of massless vector bosons) \SS. Secondly, the topology of the Nilmanifolds support fractional Wilson lines and
corresponding fractional Chern-Simons invariants, which provide useful small coefficients of the corresponding
terms in our moduli potential. This feature of our construction is similar to the strategy applied earlier to
heterotic Calabi-Yau compactifications in \fractCS. In the present case of compactifications on nilmanifolds,
there is an infinite sequence of spaces with a finer and finer discretuum of fractional Chern-Simons terms.

The topology also supports new sectors of wrapped branes. In order to introduce enough perturbative competing
forces to obtain de Sitter solutions, we introduce $KK$ monopoles (which are five-branes filling space and
wrapping a two-cycle in the compactification). These branes break the supersymmetry at a high but controllable
scale: the supersymmetry breaking scale is at the lowest of the KK mass scales in the geometry. For this reason
-- and also to exhibit the basic physical forces in the problem -- we work directly with the scalar potential in
four dimensions.
(See \DineGX\ for SUSY-breaking orbifolds of the $AdS_4$ models of \DeWolfeUU\ which also break supersymmetry at
a high scale in a controlled way.)\foot{For recent discussions of the SUSY packaging of the effective action
from compactifications on nilmanifolds with various fluxes, see e.g. \refs{\VZ,\robbins}.}

It would be interesting to investigate the possibility of a similar mechanism preserving lower-energy SUSY, and
we will mention some ideas in this direction.  In any case, it is perhaps worth emphasizing that after
supersymmetry breaking, the methods for gaining control of solutions in the effective theory are essentially the
same at different scales of supersymmetry breaking: one requires control over perturbative quantum and
$\alpha^\prime$ corrections via a well-defined approximation scheme in which the forces used to stabilize the
moduli are the dominant ones. The non-renormalization theorems of supersymmetry, while helpful in restricting
the set of corrections to compute, can at the same time complicate moduli stabilization by preventing useful
contributions to the moduli potential in the first place. For this reason we obtain our explicit solutions most
easily without imposing low-energy SUSY, although the simplifications of ten-dimensional SUSY will play a useful
role.

We will exhibit sequences of solutions which have parametrically small curvature, flux densities, and string
coupling as we take an integer quantum number $M$ to be large.  Moreover, in the solutions there is a tuneably
small ratio between the de Sitter Hubble scale $H$ and the masses of the scalar fields.

However, there is a subtlety in our construction. In the large $M$ limit, although the curvature becomes
parametrically weaker we will have a small radius $L_x\sqrt{\alpha'}\ll\sqrt{\alpha'}$ along two directions of
the solution (the circle fibers of the nilmanifolds).  Relatedly, in the simplest version of the construction,
the heaviest moduli masses end up of the same order as the lightest KK and winding masses.  Nonetheless, we will
exhibit a representative numerical solution at $M$ of order 10, for which $L_x$ is not substringy and for which
the corrections are expected to be small since the couplings and curvatures are weak. Finally, in the parametric
$M\gg 1$ limit we will suggest a more elaborate method to push the (otherwise marginally overlapping) moduli,
winding, and KK scales apart from each other (by introducing extra NS5-branes which locally reduce the inverse
string coupling and hence the KK fivebrane tensions).  In any case, this small-$L_x$ limit remains a regime of
low curvature and approximate supersymmetry as we will explain at the relevant points.

A nilmanifold by itself could simply be T-dualized along the circle fiber directions to a torus with
Neveu-Schwarz $H$ flux \KachruSK, but our construction involves other ingredients such as nontrivial
Neveu-Schwarz $B$ fields and $H$ flux and we will stay in our original T-duality frame for convenience.
It is interesting to note that in the models of \DeWolfeUU, the moduli masses were of order the curvature scale
of the $AdS_4$. Here, this problem is avoided, with the moduli masses ending up well above the Hubble scale of
our de Sitter -- but in the simplest parametric limit they bump up against the next higher scale in the problem,
the mass scale of the KK modes.


In model building in general and moduli stabilization in particular, it is important to separate the ``forest"
(the general mechanisms) from the ``trees" (the idiosyncrasies of a given construction).  One of the general
lessons of the present work -- obtained via a simple way of organizing the analysis -- is that the $AdS_4$
models \refs{\DeWolfeUU,\VilladoroCU,\CamaraDC,\GrimmUA,\GranaKF}\ and the like admit ``uplifting" terms in
their potential from a combination of negative scalar curvature and branes. At the level of the overall volume
and string coupling, the first point was also made recently in the interesting work \mit\ which we received as
this paper was in preparation (and see \SaueressigES\ for an investigation of using quantum effects to obtain dS
solutions).  A second general lesson is that the topology of spaces of negative scalar curvature naturally
supports fractional Wilson lines and fractional Chern-Simons invariants, which yield useful small coefficients
in the potential.

Although a generic cosmological solution in the landscape is quite complicated -- a fact that may be crucial for
modeling the observed scale of the dark energy as proposed in \BoussoXA\ -- specific models are useful. If
completely explicit, such constructions remove the possibility of a conspiracy working against the genericity
arguments employed in the general proposals.  Conversely, their details expose limitations to tuneability of
parameters in specific contexts.  In this spirit, recently a clean no-go result for inflation in IIA Calabi-Yau
compactifications (with a subset of the possible orientifolds, fluxes, and branes) was given in \mit.  In
constructing de Sitter in IIA, we were naturally led to ingredients going beyond the assumptions of \mit, and
together these results make it possible to focus on an appropriate set of degrees of freedom to obtain
accelerated expansion in IIA.

In particular -- and this is one of the main motivations of the present work -- explicit constructions
facilitate concrete study of the question of what microphysical degrees of freedom are required to formulate
cosmological spacetimes, perhaps in the same way that concrete black brane solutions facilitated the development
of black hole state counting and the AdS/CFT correspondence. Nilmanifolds, like hyperbolic spaces, play a
central role in geometric group theory \gromov, and hence compactifications on them may be of further conceptual
interest (either at the perturbative level \refs{\Dduality,\alex}\ or holographically \GibbonsMU).

The organization of the paper is as follows.  In \S2, after recording our ten dimensional action and conventions
we provide a convenient way of organizing the problem of checking for de Sitter minima of the moduli potential.
In \S3, we describe a particular class of models on nilmanifolds in detail. We first describe each ingredient
and its contribution to the four dimensional potential, noting subtle features as they arise.  Next, we
demonstrate the stabilization of the coupling and volumes analytically and numerically, noting the behavior of
the relevant scales in the parametric limit of interest and suggesting a more involved setup which separates the
scales further.  We separately analyze the angular moduli, showing how curvature and the other ingredients
source them and can be arranged to lift them; we also note various orbifold variants of the model which could be
used to project out many of the angular moduli. In \S4, we outline a generalization to simple de Sitter
solutions on Sol manifolds.  Finally, we conclude in \S5.  Illustrated step-by-step instructions are included at
the end.

\newsec{Preliminaries}

\subsec{Action and conventions}

We will follow some of the conventions of \DeWolfeUU\ (which itself followed \GrimmUA); for example our RR
fields satisfy $C_{RR}=C_{RR}^{Polch}/\sqrt{2}$ relative to the conventions in \PolchinskiRR. We will start from
the ten-dimensional limit of type IIA string theory, for which the action contains kinetic terms
\eqn\Skin{{\cal S}_{kin}= {1\over 2\kappa^2}\int d^{10}x\sqrt{-G}\left({e^{-2\phi}}\left({\cal
R}+4(\del_\mu\phi)^2-{1\over 2}|\tilde H|^2\right)-\left(|\tilde F_2|^2+|\tilde F_4|^2+m_0^2\right)\right)}
(with $|F_p|^2=F_{\mu_1\dots\mu_p}F^{\mu_1\dots\mu_p}/p!$ and $2\kappa^2=(2\pi)^7(\alpha')^4$). Here the full
field strengths include Chern-Simons terms
\eqn\Ffull{\eqalign{& \tilde H_3=dB+H_3^{bg}\cr & \tilde F_2=dC_1+m_0B
\cr & \tilde F_4=dC_3-C_1\wedge
H_3-{m_0\over 2}B\wedge B
}}
where the RR zero-form field strength $F_0$ is denoted $m_0$ as in \DeWolfeUU\
%
%
We will need zero-form, 3-form
and 6-form fluxes.  As discussed in \DeWolfeUU, the flux quantization
conditions can be written as
\eqn\fluxquant{m_0={f_0\over{2\sqrt{2}\pi\sqrt{\alpha'}}} ~~~~ p=(2\pi)^2\alpha' h_3
~~~~ K={1\over\sqrt{2}}(2\pi)^5(\alpha')^{5/2}f_6}
where $f_0,f_6$ and $h_3$ are integers.  Here $H=p\omega_3$ is the NS flux through a 3-cycle $\hat\Sigma_3$,
with $\omega_3$ normalized such that $\int_{\hat\Sigma_3}\omega_3=1$,
and $F_6=K\omega_6$ where the integral of $\omega_6$ over the compactification manifold is equal to one. We will
also make use of fractional Wilson lines descending from $B$ and the corresponding fractional Chern-Simons
invariants following from \Ffull\ in the presence of nontrivial $m_0$ flux. The $B$ field is normalized in the
conventional way so that it appears in the worldsheet path integral via the factor
$Exp[{i\over{2\pi\alpha'}}\int B]$.  Its periodicity is
\eqn\Bnorm{\int_{\Sigma_2}B=\int_{\Sigma_2} B+(2\pi)^2 n}
for integer $n$.
%
%

In the next section, the curvature and flux terms in \Skin\ as well as orientifold planes and KK5-branes will
yield contributions to the potential energy ${\cal U}$ in four dimensions upon compactification on a $\IZ_2$
orientifold ${\cal N}/\IZ_2$ of volume $L_0^6(\alpha')^3/2$.  (The covering space ${\cal N}$ of the orientifold
has volume $L_0^6$, hence our notation.)  We will also mention the possibility of further orbifolding
prescriptions, which would modify the volume, flux quantization, and tadpole cancellation conditions in a way
which can be obtained via a straightforward generalization of the unorbifolded case.

As reviewed explicitly in \reviews, it is most convenient to work in four-dimensional Einstein frame obtained by
scaling out of the kinetic terms the dependence on the dynamical scalars. Denoting $e^\phi\equiv g_s
e^{\tilde\phi}$ and $L^6/2\equiv (L_0^6/2) e^{6 \sigma}$, with $\tilde\phi$ and $\sigma$ fluctuating scalar
fields, we change variables to
\eqn\Eframe{G_{\mu\nu,E}^{(4)}=e^{6\sigma-2\tilde\phi}G_{\mu\nu,S}^{(4)}}
where $G_{\mu\nu,S}^{(4)}$ denotes the four-dimensional components of the string-frame metric $G$ appearing in
\Skin. The four-dimensional potential energy density in Einstein frame is then given by
\eqn\Udef{{\cal U}=M_4^4 {e^{4\phi}\over (L^{6}/2)^2}{\cal U}_s.}
where $M_4\sim {L_0^3\over g_s\sqrt{2\alpha'}}$ is the four dimensional Planck mass scale and ${\cal U}_s\equiv
-{1\over 2(\alpha')^2}\int_{{\cal N}/\IZ_2}e^{-2\phi} {\cal R}+\dots$ is the potential energy in string frame.

\subsec{Structure of the Potential}

Starting from a type II perturbative string limit and defining the $4d$ coupling
\eqn\gdefinition{g={e^\phi\over (L^3/\sqrt{2})}}
the moduli potential in four-dimensional Einstein frame has the form
\eqn\Uform{{\cal U}=M_4^4 g^2(a-bg+cg^2)+\dots}
%
where $a,b,c$ depend on other moduli $\sigma_I$. Taking the case with $a,b,c>0$ at the minimum in the $\sigma_I$
directions, and solving the quadratic equation obtained from imposing $g\del_g{\cal U}=0$ reveals \MaloneyRR\
that at fixed $a,b,c$, a positive energy solution obtains if
\eqn\dScond{1<{{4ac}\over b^2}<{9\over 8}.}
Violating the lower constraint yields AdS rather than dS, while violating the upper constraint removes the local
minimum in the potential in the $g$ direction; i.e. for $4ac/b^2=9/8$ there is an inflection point in the
potential.
Parameterizing the third coefficient in \Uform\ by $c={b^2\over{4a}}(1+\delta)$, the range \dScond\ corresponds
to $0<\delta<1/8$ and the potential takes the form
\eqn\Usmall{{{\cal U}\over M_4^4} =  g^2a\left(1-{b\over{2a}}g\right)^2+\delta{b^2\over{4a}} g^4}
Because the algebra involved in minimizing the potential can get somewhat complicated in practice, it proves
useful to organize the problem by first minimizing $\delta$ at a value slightly above zero and then showing that
there is a nearby minimum of ${\cal U}$ itself.

In particular, we will exhibit a compactification with the following property. Minimizing the quantity $4ac/b^2$
as a function of the other moduli $\sigma_I$ yields a value in the range \dScond; we will explicitly use
discrete quantum numbers to tune the minimal value of $4ac/b^2$ to be close to but slightly greater than 1. That
is, we start in a configuration $\sigma_I=\sigma_{I,0}$ minimizing $\delta$, with $\delta_0\approx 0$.

If we had $\delta_0=0$, then ${\cal U}$ would be minimized in the $g$ direction at $g_0=2a_0/b_0$, and it is
immediately clear that the potential \Usmall\ would rise quadratically in each direction in field space away
from the configuration $\sigma_{I0},g_0=2a_0/b_0$.  That is, minimizing $\delta$ would also minimize ${\cal U}$
(at fixed $g=g_0$). For $\delta_0$ tuned to be small but nonzero, there is still a local de Sitter minimum of
the potential which is close to $\sigma_{I0},g_0$ in field space, as can be seen as follows. For
$g=g_0=2a_0/b_0$ and $\sigma_I=\sigma_{I0}$ (the values minimizing $\delta$), with small positive $\delta_0$,
the potential is of order
\eqn\potvalue{{\cal U}\sim \delta_0\bar{\cal U}}
and there is a small tadpole
\eqn\smalltad{{\del{\cal U}\over{\del\sigma_I}}\sim \delta_0\bar{\cal U}}
where $\bar{\cal U}$ is of the same order as the individual terms in the potential $a_0g_0^2\sim b_0g_0^3\sim
g_0^4 {b_0^2\over 4 a_0}$.  The distance the fields are pushed by this tadpole is small, however, because in
this configuration $\sigma_{I0},g_0$ there is also a positive quadratic term which is not suppressed as
$\delta_0\to 0$:
\eqn\largequad{{\del^2{\cal U}\over{\del\sigma_I^2}} \sim {\del^2\delta\over{\del\sigma_I^2}}\bar{\cal U}\sim
\bar{\cal U} }
Similar scalings to \smalltad\largequad\ apply to the derivatives with respect to the dilaton.  The result is
that the small tadpoles shift the fields a distance of order $\delta_0$ in field space to a local minimum.  At
this local minimum, the potential is still of order \potvalue\ (plus subleading terms of order $\delta_0^2$).

After specifying our model, we will show analytically that there is a minimum $\delta_0$ of $\delta$, which can
be tuned close to zero by appropriate choices of discrete quantum numbers, thus providing a de Sitter minimum of
the potential. We will then check our results by numerically exhibiting the corresponding minimum of the
potential for the coupling and volumes, for specific values of the discrete quantum numbers.

\newsec{Models on Nilmanifolds}

Let us start from type IIA string theory in ten dimensions and consider a compactification on an orientifold of
a product ${\cal N}\equiv {\cal N}_3\times \tilde {\cal N}_3$  of two Nil three-manifolds (a.k.a. twisted tori,
a.k.a. spatial sections of Bianchi II cosmologies, a.k.a. three-tori with ``metric flux"). These manifolds are
obtained starting from the noncompact geometry
\eqn\symmmet{\eqalign{{ds^2\over\alpha'} & = L_{u_1}^2du_1^2+L_{u_2}^2du_2^2 + L_x^2\left(dx+{M\over 2}[u_1 d
u_2-u_2 d u_1]\right)^2 + \cr & + L_{u_1}^2d\tu_1^2+ L_{u_2}^2 d\tu_2^2 +  L_x^2\left(d\tx+{M\over 2}[\tu_1 d
\tu_2-\tu_2 d \tu_1]\right)^2  \cr & =
L_{u_1}^2\eta_1^2+L_{u_2}^2\eta_2^2+L_x^2\eta_3^2+L_{u_1}^2\tilde\eta_1^2+L_{u_2}^2\tilde\eta_2^2+L_x^2\tilde\eta_3^2
\cr }}
%
where $M$ is an integer and $ \eta_1=du_1, ~ \eta_2=du_2, ~ \eta_3=dx+{M\over 2}[u_1 d u_2-u_2 d u_1]$ are
one-forms invariant under the Heisenberg group of isometries of the nilgeometry (and similarly for the tilded
coordinates). We compactify this space by making identifications on the coordinates by a discrete subgroup of
the isometry group generated by elements:
\eqn\compactification{\eqalign{& t_x: ~~ (x,u_1,u_2, \tx,\tu_1,\tu_2) \to (x+1, u_1,u_2, \tx,\tu_1,\tu_2) \cr &
t_1: ~~(x,u_1,u_2, \tx,\tu_1,\tu_2) \to (x-{M\over 2}u_2, u_1+1,u_2,\tx,\tu_1,\tu_2)\cr & t_2: ~~ (x,u_1,u_2,
\tx,\tu_1,\tu_2) \to (x+{M\over 2}u_1,u_1,u_2+1,\tx,\tu_1,\tu_2) \cr}}
and similarly for the tilded coordinates.  The Nil 3-manifold can be described as follows.  For each $u_1$,
there is a torus in the $u_2$ and $x'\equiv x-{M\over 2}u_1u_2$ directions (under this change of coordinates we
have $\eta^3=dx'+Mu_1du_2$). Moving along the $u_1$ direction, the complex structure $\tau$ of this torus goes
from $\tau\to \tau +M$ as $u_1\to u_1+1$. The projection by $t_{u_1}$ identifies these equivalent tori.  The
directions $u_1$ and $u_2$ are on the same footing; similar statements apply with the two interchanged and with
$x'$ replaced by $x''\equiv x+{M\over 2}u_1u_2$.

We will orientifold the space by an exchange of the tilded and untilded coordinates combined with an exchange of
left and right movers; hence the volume of our compactification will be $L^6/2$ where
\eqn\volfix{L^6=L_x^2L_{u_1}^2L_{u_2}^2}
is the volume of the full compact space ${\cal N}$ in string units.

The projections \compactification\ generate the fundamental group of ${\cal N}$.  They satisfy the relation
\eqn\pione{t_2t_1t_2^{-1}t_1^{-1}=t_x^M}
The first homology is given by the abelianization of the fundamental group, obtained by setting all commutators
to the identity.  For each nil three-manifold this is $\IZ^2\times\IZ_M$. The last factor comes from cycles
introduced by the projections $t_x^m$ with $m<M$ (since for $m<M$ these elements are not commutators in the
fundamental group, and hence are not set to the identity by the abelianization). The nilmanifold with $M\ne 1$
is a freely acting $\IZ_M$ orbifold of the a nilmanifold with $M=1$, obtained by projecting by translations
along the $x$ direction. The smaller cohomology group than that of the 3-torus arises because of the relation
\eqn\homreln{d\eta_3=M\eta_1\wedge\eta_2,}
which means $\eta_3$ is not closed and $\eta_1\wedge\eta_2$ is exact, reducing by one the dimension of $H^1$ and
$H^2$. As a result, there are fewer continuous moduli from NS and RR gauge potentials on nilmanifold
compactifications as compared to tori, and there are additional vacua corresponding to discrete Wilson lines
which we will employ.

%
%
%
%
%

The compact nilmanifold also has a reduced isometry group: upon compactification \compactification, the
nilmanifold retains only the $U(1)$ isometry corresponding to continuous shifts of $x$, in contrast to the
$U(1)^3$ isometry group of $T^3$. As mentioned above, this will help lift some of the scalar degrees of freedom
which are eaten in the generalized Higgs mechanism explained in \SS.\foot{This fact that a non-Ricci-flat
compactification introduces fewer moduli than its Ricci-flat counterpart is an example of a more general
phenomenon; hyperbolic spaces of dimension greater than two are famously rigid, there being no continuous
deformations of the isometry groups used to compactify the space by projection from the hyperboloid.}

We have chosen a symmetric configuration \symmmet\ to expand around. This renders the analysis simpler since
enhanced symmetry points are automatically extrema of the full effective potential in symmetry-breaking
directions.  Of course we must lift all the light scalar fields including the symmetry-breaking approximate
moduli of the metric as well as $L_{u_1},L_{u_2},L_x$ and the string coupling $g_s$. Throughout the
construction, for simplicity we will maintain a symmetry between $(x,u_1,u_2)$ and $(\tx,\tu_1,\tu_2)$, a
symmetry which will be enforced by an orientifold projection.

%
%
%
%

The scalar curvature of ${\cal N}={\cal N}_3\times \tilde{\cal N}_3$ is
\eqn\Rscalar{{\cal R}=- {L_x^2M^2\over \alpha' L_u^4}}
where
$L_u^4\equiv L_{u_1}^2L_{u_2}^2$ This contributes a positive term in the
four-dimensional Einstein frame potential energy \Udef\
\eqn\potR{{\cal U}_{{\cal R}}= {M_4^4\over 2} {L_x^2M^2\over L_u^4} {e^{2\phi}\over (L^6/2)} = {M_4^4\over 2}
{e^{2\phi}M^2\over (L^{6}/2)}{L_x^4\over L^6}=M_4^4 {g^2L_x^4M^2\over{2 L^6}}}
descending from the $10d$ Einstein-Hilbert action, where we used \volfix\ and
%
%
the above definition of $g$ \gdefinition.  An important feature of \Rscalar\potR\ is the fact that for $M
L_x/L_{u_1}\ll 1$ and $M L_x/L_{u_2} \ll 1$, the inverse curvature radius is smaller than the KK scales
$1/L_x,1/L_{u_1},$ and $1/L_{u_2}$.  This is related to the fact that the nilmanifold is T-dual, along the $x$
direction, to a $T^3$ with NS three-form flux -- a system for which moduli masses are below the KK scale of the
$T^3$ \KachruSK.  As the $x$ circle {\it shrinks} ($L_x\to 0$) the curvature becomes {\it weaker}.


In our final solution, the curvature and flux densities will be small\foot{For substantial recent progress
controlling worldsheet theories with substantial curvature and $H$ flux, see \Allan.} but $M$ will be large
enough that $L_x^2M\sim 1$ parametrically as $M\to\infty$. At modest values of $M$ (e.g. $M\sim 10$), we will
find numerically that $L_x$ can be slightly greater than string scale. In the parametric limit at small $L_x$,
one can consider T-dualizing to obtain a large circle, but we will continue to describe the system in the
original T-duality frame. One reason for this is that our solution will involve discrete Wilson lines from the
NS $B$ field as well as NS flux, which complicate the T-duality transformation.
Although $L_x$ gets small in this limit, the winding modes will remain parametrically at least as heavy as the
moduli and the lightest KK modes in all versions of the constructions.


In the most symmetric case where the lightest KK modes are not be parametrically heavier than the heaviest
moduli, but we will still analyze and stabilize the moduli fields separately from the KK modes, for two reasons.
First, the KK (and winding) modes --treated separately themselves -- are massive (those with winding or momentum
on the $T^2_{x,\tx}$ exhibiting interesting Landau level degeneracies on the nilmanifold with $H$ flux \alex).
This together with our analysis of the moduli masses will establish that the diagonal blocks in the moduli and
KK mass matrices are positive. The main remaining question in this version of the construction is then whether
large off-diagonal terms in the mass matrix could arise. As we will see after assembling our ingredients, the
symmetries of the problem help suppress mixing between the lightest KK modes and the heaviest moduli, suggesting
that this need not happen (though we have not analyzed this combined problem in nearly as much detail as the
moduli themselves).  Because of this uncertainty, we will also suggest a generalization of the model with an
extra ingredient which allows us to push the marginally overlapping scales apart.


This contribution \Rscalar\potR\ pertains to the diagonal metric \symmmet; the curvature will also depend on a
subset of the off-diagonal deformations, lifting them in a way originally computed in \SS.  The approximate
moduli of nilmanifolds were laid out in a form respecting the symmetry structure of the theory in \SS. They
consist of metric deformations, deformations of the NS NS two-form potential, and RR axions.

Let us start with the metric moduli.   The metric modes are
\eqn\metmod{\Delta ds^2={\cal G}_{IJ}\eta^I\eta^J+{\cal G}_{\tilde I\tilde J}\tilde\eta^{\tilde
I}\tilde\eta^{\tilde J}+{\cal G}_{I\tilde J}\eta^I\tilde\eta^{\tilde J}}
with $G_{IJ}\equiv G_{\tilde I\tilde J}$ enforced by an orientifold action we will introduce below. Of these,
${\cal G}_{xu_i},{\cal G}_{\tx\tu_i}$, and ${\cal G}_{u_i\tilde u_j}$ are lifted in the Higgs mechanism
explained in \SS. As discussed above, in contrast to the $U(1)^3$ isometry group of a $T^3$, only one $U(1)$
isometry $x\to x+\lambda$ survives from each Nil three-manifold.  In compactification, isometries yields
lower-dimensional gauge bosons from off-diagonal metric modes. The would-be gauge bosons corresponding to the
broken $U(1)^2$ still exist in the present case of a twisted three-torus, but in a Higgsed phase.

Consider now the ${\cal G}_{x\tx}$ mode. Dimensionally reducing first on one of the nilmanifold factors, say
${\cal N}_3$, this is a component $A^{met}_{\tx}$ of a $U(1)$ gauge boson $A^{met}_\mu$ arising from the
continuous isometry of the metric in the $x$ direction.  The Wilson line of this gauge boson around the $\tx$
direction is constrained by the relation \pione:
\eqn\metWL{(e^{i\oint_{\gamma_{\tx}} A^{met}})^M=1 \Rightarrow \oint_{\gamma_{\tx}} A^{met}=2\pi{q'\over M} ~~~
q'\in \IZ.}
In particular, there is not a continuous Wilson line degree of freedom associated with this mode:  the mode
$A_{\tilde 3}\tilde\eta^3$ is massive, as can be seen from the formulae in \SS\ (where the discrete Wilson line
degree of freedom is not described directly). A similar statement holds also for the NS $B$ field. We will find
that discrete Wilson line degrees of freedom are very useful in our setup, and will describe them and their
effects in more detail below.

Among the angular moduli, this curvature potential leaves unfixed the modes ${\cal G}_{u_1u_2},{\cal
G}_{\tu_1\tu_2},{\cal G}_{u^i\tx}={\cal G}_{x\tu^i}$ ($i=1,2$).  The modes which are lifted by the curvature can
also get contributions from other terms in the moduli potential, as we will discuss in analyzing the angular
moduli below in \S3.7.
In addition, we must stabilize the diagonal moduli ${\cal G}_{II}={\cal G}_{\tilde I\tilde I}$, equivalently
$L_x^2,L_{u_1}^2,L_{u_2}^2$ (one of which can be traded for the overall volume mode \volfix, which is a runaway
direction in field space).

Some of the $B, C_1,C_3$, and $C_5$ fields will be lifted by a combination of the $\tilde F^2$ terms from
\Skin\Ffull\ and by the orientifold projection (and in more general examples, orbifold projections).  Others
will be unsourced by the leading terms in the potential, and be fixed by higher order, lower scale effects.

\subsec{Orientifold and fluxes}

In order to obtain a metastable solution, we will require a negative term in the potential, at an intermediate
order in the expansion about weak coupling and large volume, since the potential energy in four dimensional
Einstein frame decays to zero at weak coupling and low curvature \reviews. To this end, introduce an O6-plane as
in \DeWolfeUU, as follows.\foot{There are other, discretely distinct, options for defining the space group of
the orientifold.  Another example would be to mod by the same orientifold action, but -- in compactifying the
original space via the discrete isometries \compactification\ -- to project only by the group generated by
elements of the form $t_yt_{\tilde y}$ and $t_{y}t_{\tilde y}^{-1}$. This would yield $2^3$ different O6-planes
from the fixed points of the group action, on a space of volume $4 L^6$.  We expect similar results for all
these cases, but the detailed factors in the potential would differ in different examples.} Mod out the
worldsheet sigma model by an exchange of tilded and untilded embedding coordinates (and Fermi partners) combined
with an exchange of left and right movers:
\eqn\Oplane{\Omega : ~~ (x,u_j,\tx,\tu_k)\leftrightarrow (\tx,\tu_j,x,u_k) ~~~~ L\leftrightarrow R ~~~~
(-1)^{F_L}}
As reviewed in \DeWolfeUU, under the orientifold transformation, $B, C_1,$ and $C_5$ are are odd. Geometrically,
\Oplane\ introduces on O6 plane wrapped on the 3-cycle traced out by the fixed point locus
$(x,u_j)=(\tx,\tu_j)$. The negative tension of the O6 plane leads to the potential energy contribution
\eqn\Otension{{\cal U}_O=-\kappa^2(2\mu_6){g_s^3\over (L^6/2)^2}(Vol_{O6})=-2^3\pi g^3}
where $\mu_6$ is the D6-brane tension (equal to $\pi/\kappa^2$) and $Vol_{O6}$ is the volume of the cycle
wrapped by the O6-plane (which in our case is $(\sqrt{2})^3L^3$).

We must cancel the O6-plane's charge.  Following \DeWolfeUU, we can use fluxes to cancel the tadpole for $C_7$.
The O6-plane constitutes a localized source of $F_2$ within the 3-cycles dual to the cohomology classes
$\eta^1\wedge\eta^2\wedge\eta^3-\tilde\eta^{1}\wedge\tilde\eta^{2}\wedge\tilde\eta^{3},
\eta^1\wedge\tilde\eta^{2}\wedge\eta^3-\tilde\eta^{1}\wedge\eta^{2}\wedge\tilde\eta^{3}$, and
$\tilde\eta^{1}\wedge\eta^{2}\wedge\eta^3- \eta^{1}\wedge\tilde\eta^{2}\wedge\tilde\eta^{3}$. The tadpole
cancellation condition is that for each 3-cycle $\Sigma_3$,
\eqn\tadpole{m_0\int_{\Sigma_3}H=-2\sqrt{2}\mu_6\kappa^2 n_{O6}
}
where $n_{O6}$ is the net number of O6-planes sitting at points in $\Sigma_3$.  Writing
\eqn\Ocycle{\eqalign{ H\equiv & p_1\left(\eta^1\wedge\eta^2\wedge\eta^3-\tilde\eta^{\tilde
1}\wedge\tilde\eta^{\tilde 2}\wedge\tilde\eta^{\tilde 3}\right)+p_2\left(\tilde\eta^{\tilde
1}\wedge\eta^{2}\wedge\tilde\eta^{\tilde 3}-\eta^1\wedge\tilde\eta^{\tilde 2}\wedge\eta^3\right)\cr &
+p_3\left(\eta^{1}\wedge\tilde\eta^{\tilde 2}\wedge\tilde\eta^{\tilde 3}- \tilde\eta^{\tilde
1}\wedge\eta^{2}\wedge\eta^3\right)}}
where $p_i\equiv -h_{3,i}(2\pi)^2\alpha'$ (c.f. \fluxquant), and imposing \tadpole\ (with $n_{O6}=1$ O6-planes
passing through each cycle) sets $f_0h_{3,i}=2$. We therefore take $h_{3i}=h_3$ for $i=1,\dots,3$, with
\eqn\Hfzero{f_0 = 1 ~~~~~ h_3 = 2}
Note that this satisfies the flux quantization condition on both the covering space and the orientifold itself.
In evaluating \tadpole, we took into account the fact that each of the 3-cycles is halved in volume by the
action of the orientifold (c.f. \FreyPol).

The O6-plane, $H_3$ flux, and $F_0$ flux together contribute the following terms to the four-dimensional
effective potential in $4d$ Einstein frame \Udef:
\eqn\potOHzero{{\cal U}_{OHm_0} = {M_4^4}\left({3 p^2 g^2\over{2(\alpha')^2
L^6}}-{2\sqrt{{2\over\alpha'}}}|m_0p|g^3+{\alpha'm_0^2g^4L^6\over 4}\right)
}

%
%
%
%
%
%

We will also include six-form flux
\eqn\sixform{F_6=K\eta^1\wedge\eta^2\wedge\eta^3\wedge\tilde\eta^1\wedge\tilde\eta^2\wedge\tilde\eta^3}
where $K=f_6(2\pi)^5(\alpha')^{5/2}/\sqrt{2}$ in terms of the integer flux quantum number $f_6$.  This leads to
the following term in the moduli potential:
\eqn\UFsix{{\cal U}_{F_6}= M_4^4 g^4 {K^2\over {4L^6(\alpha')^5}}}
%


%
%


\subsec{Fractional Chern-Simons invariants}

The $\IZ_M\times\IZ_M$ homology cycles described above yield a set of discrete Wilson line vacua, and
corresponding fractional Chern-Simons invariants which we will use to obtain contributions to the effective
potential with tuneably small coefficients, as follows.\foot{See \fractCS\ for a previous example using
fractional CS invariants to help in heterotic moduli stabilization.} As discussed in \S2.1, the effective action
contains a term $-{1\over 2\kappa^2}\int d^{10} x |\tilde F_2|^2$ where
\eqn\Ftwofull{\tilde F_2=dC_{(1)}+m_0B
}
with $m_0$ the RR 0-form flux \fluxquant\ and $B$ a Neveu-Schwarz two-form potential.  In our background
solution, $B$ will be flat (note that we separated the $H$ flux from $dB$ in \Ffull\ as reviewed in \DeWolfeUU).
Its fractional Wilson line vacua lead to fractional Chern-Simons forms $m_0B$, which will provide useful small
coefficients in the moduli potential.


In general, a manifold with nontrivial fundamental group $\pi_1$, can support discrete Wilson line vacua of
gauge fields of a gauge group $G$ -- flat connections with nontrivial holonomy around non-contractible cycles.
As reviewed in \GSW, they correspond to homomorphisms from $\pi_1$ into $G$, since Wilson lines $U_\gamma=P
Exp(i\int_\gamma A)$ must satisfy the group multiplication law $U_{\gamma\gamma'}=U_\gamma U_{\gamma'}$. Since
$G$ is abelian in our case, the only discrete Wilson lines arise from closed paths which are nontrivial in
homology (which is the abelianization of the fundamental group obtained by setting commutators to 1); elements
$g$ of $\pi_1$ which are commutators (elements of the form $g=g_1g_2g_1^{-1}g_2^{-1}$) have trivial holonomy
$U_g=1$. The first homology group of ${\cal N}_3$ includes the $\IZ_M$ factor, represented by the closed paths
$\gamma_m$ introduced by the projection $t_x^m$ with $m<M$ \compactification.
%
%

%
%
%
%
%

%
%

Flat connections for gauge fields on nilmanifolds were derived explicitly in \WLfiber. This construction
generalizes to the NS 2-form potential. In a local neighborhood of ${\cal N}$, and for $0\le x<1,0\le \tx<1$, we
can take $B$ to be
\eqn\B{\eqalign{B=& {q\over M}(2\pi)^2\alpha'dx\wedge d\tx +{r\over
M}(2\pi)^2\alpha'\left(dx\wedge\tilde\eta^1-d\tx \wedge\eta^1 \right)+(1\leftrightarrow 2) +\dots
}}
The first terms of \B\ contain the $B$ field analogue of discrete Wilson lines.  These are projected in by the
orientifold action \Oplane\ (using the fact reviewed in \DeWolfeUU\ that $B$ has an intrinsic parity under the
orientifold, c.f. eqn (2.9) of \DeWolfeUU). The rest indicated by $\dots$ contains the continuous Wilson lines
invariant under the orientifold; these are lifted by the $|\tilde F_2|^2$ term in the potential. The form for
$B$ in different neighborhoods is then derived by using the transition functions between them as in \WLfiber.

The discrete Wilson line terms in \B\ yield
\eqn\BWL{e^{{i\over{2\pi\alpha'}}\int_{\gamma_x\times\gamma_{\tx}}B}=e^{2\pi i {q\over M}}
 ~~~~~~~~
e^{{i\over{2\pi\alpha'}}(\int_{\gamma_x\times\gamma_{\tu_1}}B-\int_{\gamma_{\tx}\times\gamma_{u_1}}B)}=e^{2\pi i
{2r\over M}} }
and introduces the potential term (from the $|\tilde F_2|^2$ and $|\tilde F_4|^2$
 terms in \Skin)
\eqn\UBWL{{\cal U}_{BWL}= 4\pi^4M_4^4{m_0^2\alpha'}\left({q\over M}\right)^2g^4{L^6\over L_x^4} + 16\pi^4 M_4^4
m_0^2\left({r\over M}\right)^2{g^4 L^3\over{L_x}}+2^8\pi^8 M_4^4 m_0^2\left({r\over M}\right)^4{g^4 \over{
L_x^2}} }
We will ultimately choose $q$ to be of order 1, and can use the ratio $r/M$ to help tune the cosmological
constant, as well as to help stabilize some angular moduli (though in that regard, orbifold variants of the
construction which remove angular moduli could also project out the terms proportional to $r/M$ -- this would be
a consistent choice, since as we will see these terms are not crucial for stabilizing the coupling and volume
moduli). In writing \UBWL, we took $L_1=L_2$, and will consider other ingredients which respect this symmetry
and consistently stabilize the system at this point.

Another way to describe the contribution in the first term of \B\BWL\UBWL\ is as an example of discrete torsion
\DT. The torsion cycles $\gamma_1$ and $\gamma_{\tilde 1}$ in our compactification manifold are obtained by the
projection $\IZ_M\times \IZ_M$ starting from a finite cover (the same space with $M=1$).  With the $B$ field \B\
turned on, the projection in the $\IZ_M$ winding string sectors are modified by the factor
$Exp[{i\over{2\pi\alpha'}}\int B]$ in the worldsheet path integral. This example has the interesting feature
that the discrete torsion is not associated with an orbifold singularity, since the projection is freely
acting.\foot{For recent studied of discrete torsion, see e.g. \SharpeWU.}

\subsec{$KK5$ branes}

The nontrivial topology of our compactification manifold can also support wrapped branes.  Spacefilling KK
monopoles \KKmon\ will play a useful role, providing a needed independent ``uplifting" term in the potential.
These are objects magnetically charged under a linear combination of the $U(1)$ isometries along the $x,\tx$
directions, and are extended in $4d$ as well as along two internal directions.  They are T-dual to $NS5$-branes
\KKNS, and we will refer to them as KK fivebranes.

As a specific example, we introduce the following set of KK5-branes.  Start with $n_K$ KK fivebranes
magnetically charged under $U(1)_x\times U(1)_{\tx}$ with charges (1,1). (These are T-dual to NS5-branes at
points on a circle of radius $\sim 1/L_x$.)  Wrap these KK fivebranes along the the transverse direction from
$x,\tx=0$ to $t_x t_{\tx}^{-1}$ and from $u_1,\tu_1=0,u_2,\tu_2=0$ to $t_{u_1}t_{\tu_1}t_{u_2}t_{\tu_2}$.  The
latter cycle is subject to ordering ambiguities because of the relation \pione, but any order will give similar
scalings in our moduli potential.\foot{As with the other ingredients, there are variants of this configuration
which could also be considered, with sets of KK5-brane stretched in various different directions.  This is
important for example in versions in which one orbifolds the geometry, in which the KK5-brane configuration
would need to be invariant under the corresponding symmetry. This can be arranged by using sets of fivebranes
respecting the orbifold symmetry, or if necessary adding other sets rotated appropriately relative to the
original set. Also, NS fivebranes which play a similar role can be used.}

\noindent{\it Topological Consistency Conditions and $n_K$}

We must make sure our configuration is consistent with all of the previous ingredients, and we must cancel all
the relevant charges within the compactification. The number $n_K$ of $KK5$-branes may be constrained to be a
multiple of $M$ or to be combined with antibranes in order to accomplish this. One reason is charge
cancellation. Each of the $n_K$ branes wrap an $\IZ_M$ homology cycle. As such, it is not a source for gauge
bosons of a continuous gauge symmetry group, so Gauss' law does not directly apply to impose charge
cancellation. However, for D-branes in this kind of situation, one does find that K-theory charges must be
cancelled in compact manifolds \Ktheory, and a similar constraint may arise in the present case.  To be safe, we
will assume such a condition holds in our present context.

Moreover, in discussing the flat connection for $B$ \B\BWL\ above, we used the fact that our compactification
manifold is a freely acting $\IZ_M\times \IZ_M$ orbifold of a nilmanifold, giving torsion 1-cycles
$\gamma_1,\gamma_{\tilde 1}$ in the $x,\tx$ directions which led to the possibility of discrete torsion \BWL. In
the presence of $n_K<M$ KK monopoles, however, strings can only be conserved mod $n_K$ \KKNS, so the $\IZ_M$
winding charge for strings wound around these cycles is no longer conserved; the cycles $\gamma_1^{n_K}$ and
$\gamma_{\tilde 1}^{n_K}$ bound 2-cycles. By Stokes' law, in this situation the fractional Wilson line ${1\over
{2\pi\alpha'}}\int_{\gamma_1\times\gamma_{\tilde 1}} B$ is quantized in units of $1/n_K$ rather than $1/M$. For
$n_K$ a multiple of $M$, with all the KK5-branes sitting at the same point in their transverse directions, we
recover the $\IZ_M\times\IZ_M$ symmetry and the consequent discrete Wilson line (discrete torsion) taken in
\B\BWL.

There are in general further topological constraints on combinations of branes and fluxes.  A canonical example
of this type of consistency condition is that of \FreedWitten :  Dp-branes with $H$ flux on their worldvolume
must have a corresponding number of $D(p-2)$ branes ending on them.  This type of condition has been generalized
to KK monopoles and NS5-branes in \VZ. For KK 5-branes, Villadoro and Zwirner \VZ\ find -- using various
U-duality arguments -- constraints on $\tilde F_2$ flux along a two-cycle consisting of the fiber circle times a
one-cycle in the brane worldvolume.  In our setup, if $n_K<M$ so that the brane wraps a homologically nontrivial
cycle in the $x,\tx$ directions, then the fractional Chern-Simons invariant $m_0B$ coming from \B\ (whose flux
quantum number is $q/M$) is nontrivial.  As far as we can tell, more analysis would be required to determine if
this leads to an anomaly, and if so whether that anomaly could be cancelled by the addition of other branes.


Because of these subtleties, we will keep track of the $n_K$ dependence but will focus on the cases where $n_K$
is a multiple of $M$, so that each set of branes is homologous to a single brane wrapping a homologically
trivial (but homotopically nontrivial) cycle.
This evades both generalized K-theoretic subtleties just listed, and also does not require additional antibranes
to cancel the charges (though these could be included).

In this case, the discrete torsion \BWL\ is consistent with the KK5-branes, as long as they are placed together.
(In the T-dual description, the NS5-branes are arranged symmetrically along the T-dual transverse circle,
restoring the $\IZ_M$ translation symmetry in that description, but are together in the remaining transverse
directions.)  This provides another example of discrete torsion helping to stabilize moduli \VafaWitten,
significantly simplifying the problem of stabilizing the 5-brane positions since their relative motion is
projected out.



\noindent{\it Potential Contribution}

For values of $L_u\gg M L_x$ for which the fivebranes are well localized within the transverse $u,\tilde u$ and
T-dual $x+\tx$ directions, they are locally supersymmetric. The BPS formulas for the tensions yield the
following contribution to the potential ${\cal U}$ from $n_K$ such sets of KK 5-branes:
\eqn\Ufive{{\cal U}_{KK5}
=M_4^4 {2 \sqrt{2 ({L_1\over L_2}+{L_2\over L_1})} {n_{K}\over\sqrt{\eta}}g^2{L_x^{5/2}\over{L^{9/2}}}}}
%
where we defined
\eqn\defeta{\eta\sim {L_{x-\tx}\over L_{x+\tx}}}
This degree of freedom $\eta$ is related to the angular metric discrete Wilson line degree of freedom describe
in \metWL.  Starting from a given discrete Wilson line vacuum, varying the continuous modulus ${\cal G}_{3\bar
3}$ by changing the angle $\gamma$ between the $x,\tx$  directions (at fixed volume) changes the lengths of the
cycles generated by $t_xt_{\tx}$ and $t_xt_{\tx}^{-1}$.  More precisely, starting from $q'=0$, in terms of the
angle $\gamma$, the ratio $\eta$ is given by
\eqn\gammaeta{\eta=\sqrt{{1-\gamma}\over{1+\gamma}}}
%

In analyzing the angular dependence of our potential terms below in \S3.7, we will require their dependence on
the
\defeta, as well as on other similar angular moduli.
In addition to the KK fivebrane contribution, the curvature, O6 and $H, F_2,F_4$ flux contributions depend on
$\eta$, reducing to \potR\Otension\potOHzero\ when $\eta\to 1$.  Since the Wilson line is discrete, and the
corresponding ${\cal G}_{3\tilde 3}$ deformation is massive, the curvature potential acquires a factor of the
form $1+(\eta-\eta_{q'})^2+\dots$. The negative O6-plane tension acquires a factor of $1/\sqrt{\eta}$, since its
length increases when $\eta$ decreases.  The $H$ flux term, which threads the dual cycle to that wrapped by the
O6-plane, scales like $1/\eta$, since the flux lives in a larger cycle when $\eta$ increases.  Similarly, the
second contribution in \UBWL\ acquires a factor of $1/\eta$ at small $\eta$.


We should make one further comment about the formula \Ufive. Since they are all together, our $M$ branes have
substantial a throat cross section (the size of each KK5-brane being given by the size $L_x$ of its fiber
direction \KKmon), which with $M$ of them adds up to a size $ML_x$. We should compare this to the size of the
compactification in the directions transverse to them.  In our simplest solution, both will be of order
$M^{1/2}$, so that the KK fivebrane cores bump up against the size of the compactification. This, along with the
ratio of KK to moduli masses, motivates a more elaborate setup separating these scales, and indeed we will
ultimately find a method to push the transverse size of the space larger than this.  In any case, \Ufive\ gives
a good estimate for the parametric scaling of the KK fivebrane contribution to the effective action. We will
return to discuss the angular and fivebrane moduli after addressing the problem of stabilizing the runaway
moduli $g,L_u,L_x$.



\subsec{Stabilization of volumes and coupling}

Altogether, we have a potential energy for $g, L, L_x$ of the form
\eqn\Ufull{\eqalign{& { {\cal U}\over M_4^4}  =ag^2-bg^3+cg^4\cr &= g^2\left(M^2 {L_x^4\over {2 L^6}}+(4
{n_{K}}) {L_x^{5/2}\over L^{9/2}}+{3 p^2\over {2(\alpha')^2 L^6}}\right)-g^3\left(2{\sqrt{2\over { \alpha'}}}|p
m_0|\right)\cr & +g^4\alpha'\left({m_0^2\over 4}L^6+4\pi^4m_0^2\left({q\over M}\right)^2{L^6\over L_x^4}
+\left({r\over M}\right)^2{16\pi^4m_0^2L^3\over{L_x}}+\left({r\over M}\right)^4{2^8\pi^8 m_0^2\over{ L_x^2}}
+{K^2\over{4 L^6(\alpha')^6}}\right)\cr }}
%
where we set $L_1/L_2=1$, since this is where it is stabilized by \Ufive, and where we suppressed dependence on
angular moduli to be discussed in the next subsection.
Following the discussion in \S2, we will proceed to show that
for suitable choices of discrete quantum numbers, the quantity
\eqn\combo{\eqalign{& {4ac\over b^2} \equiv 1+\delta(L,L_x) \cr & ={(\alpha')^2\over{2 p^2m_0^2}}\left(M^2
{L_x^4\over {2 L^6}}+(4 {n_K}) {L_x^{5/2}\over L^{9/2}}+{3 p^2\over {2(\alpha')^2 L^6}}\right)\cr
&\times\left({m_0^2\over 4}L^6+4\pi^4m_0^2\left({q\over M}\right)^2{L^6\over L_x^4}
+\left({r\over M}\right)^2{16\pi^4 m_0^2 L^3\over{L_x}}+\left({r\over M}\right)^4{2^8\pi^8 m_0^2\over{ L_x^2}}
+{K^2\over{4 L^6(\alpha')^6}}\right)\cr}}
has a minimum $L_0,L_{x0}$ in the space of $L,L_x$ which is in the range
\eqn\window{0<\delta(L_0,L_{x0})<{1\over 8}}
with $\delta(L_0,L_{x0})$ tuneable to be small.

We need to make sure first of all that the minimum of \combo\ is not above 9/8, which we can show as follows.
First, note that at \combo\ includes a constant term $q^2/(16 h_3^2)+3/16$ (from the $1st\times 2nd$ and $3rd
\times 1st$ cross terms, respectively).

Next, consider the two terms (from the $1st\times 1st$ and $3rd\times 2nd$ cross terms, respectively)
\eqn\twonext{{\alpha'\over{2 p^2m_0^2}}\left({M^2m_0^2\alpha'\over 8}L_x^4+{6\pi^4
m_0^2p^2q^2\over{\alpha'L_x^4M^2}}
\right)}
The terms \twonext\ are minimized at $q\sqrt{3}\over{8h_3}$, with
\eqn\firstmin{L_x^4\sim {1\over M^2}
}

At fixed $L_x$, the remaining terms in \combo\ diverge for $L\to 0$ and for $L\to\infty$.
Thus $4ac/b^2$ has a minimum at finite
nonzero values of $L,L_x$.   We will next show that this minimum is tuneable to lie in the range \dScond\ as
close as desired to the lower limit as $M\to\infty$, and show that the corrections to our solution are small. We
will then verify numerically that a de Sitter minimum of the potential \Ufull\ exists.

Plugging \firstmin\ into \combo, it reduces to $(3/16)+(q^2/(16 h_3^2))+{q\sqrt{3}\over {8h_3}}$ plus a function
of $L$ which diverges as $L\to 0$ or $L\to\infty$.
The fact that it diverges as $L\to 0$ for any value of $r$ (including zero) is a consequence of the six-form
flux contribution \UFsix; in the absence of this contribution, the minimum of $4ac/b^2$ would be at $L=0$, with
the value $(3/4)+(q^2/(4 h_3^2))+{q\sqrt{3}\over {2h_3}}$. In the presence of \UFsix, the minimum can therefore
be tuned to sit within the required range \dScond\ by adjusting $q$, $r$, $M$ and $K$.

We will also keep track of factors of $n_K/M$ even though we will consider the case $n_K= M$ for the topological
reasons discussed above.  The reason for this is that the resulting formulas will make clear that in order to
generalize the construction to separate its mass scales further, it would be useful to introduce effects which
decrease the contribution of the KK fivebranes.  We will ultimately propose to do this (below in \S3.6) not by
reducing the number $n_K$ of them, but by reducing their tension by placing them in a local region the
compactification with an enhanced string coupling.

Now consider the two terms
\eqn\twofirst{{(4 n_K)(\alpha')^2L_x^{5/2}\over{2 p^2m_0^2}}\left({m_0^2\over 4}L^{3/2}
+{K^2\over{4L^{21/2}(\alpha')^6}} \right)}
These terms are minimized at a value of order ${ n_K\over M}({K\over M})^{1/4}$
obtained for $L\sim K^{1/6}$.
Taking into account that $L_x$ is constrained by \twonext, this term combined with the first term of \twofirst\
prevents any decay mode with $L\to\infty$ or $L\to 0$.

In general all the terms in \combo\ are consistently of the same order at the minimum, with the parametric
scalings
\eqn\newscales{\left( {n_K\over M}\right)\sim \left({M\over K}\right)^{1/4}
~~~~ L\sim K^{1/6} ~~~~ L_x\sim {1\over M^{1/2}}}
As emphasized above, $L_x$ ends up small in our parametric limit \firstmin.  For finite values of $M$, of
course, the results depend on order 1 factors.  Below, in \S3.8, we will exhibit a numerical local minimum of
the potential \Ufull\ in the $g,L,L_x$ directions for which $L_x$ ends up $\sim 2$ at $M=10$.

\subsec{Scales and Mixing}

%
%


Using \newscales\ we can now indicate the physical scales of interest in our solution (again assuming that as
discussed below in \S3.7, the angular moduli stay near the original point \symmmet\ about which we have
expanded).
The string coupling $g_s = g L^3$ is
\eqn\gstring{g_s\sim {2a\over b}L^3 \sim {1\over L^3}\sim
{1\over K^{1/2}}}
Since we have a small cycle $L_x$, it is interesting to consider the T-dual coupling $\hat g_s$ (even though the
T-duality affects the ingredients listed above in a somewhat complicated way).  This is
\eqn\gdual{\hat g_s^2\sim {g_s^2\over L_x^4}\sim {M^2\over K}\sim M\left({n_K\over M}\right)^4}
Thus in our simplest setup with $n_K=M$, this is large, but if we can elaborate the model to lower the
contribution of the KK fivebrane configuration, this could be small; we will suggest a method for achieving this
in the next subsection.  Since the theory is approximately supersymmetric, the corrections are parametrically at
most of order $\hat g_s^2{\cal R}\sim \hat g_s^2/L^6\sim (M/K)^2\ll 1$.  In the numerical solution of \S3.8\ at
a modest value for $M$ of 10, we will see that both $g_s$ and $\hat g_s$ can remain $\le$ order 1 with small
curvatures, leading to suppressed corrections.  Even in the parametric $M\to\infty$ limit of the simplest
version of the construction, the estimate just given might be too pessimistic:  if in the T-dual model one
crossed over into the M theoretic regime with $\hat g_s\gg 1$, the corrections to the moduli potential should
not grow with increasing size of the eleventh dimension, but should fall off at large radius.

Another scale of interest is the core size of our set of KK 5-branes as compared to the size of the
compactification. This is of order $M^{1/2}$, and again will bump up against the size of the compactification
$L_u\sim (KM)^{1/4}$ in the simplest version, again motivating increasing the ratio $K/M$.

Using the fact that the canonically normalized moduli fields are the logs of $g_s,L,L_x$ we have that the scale
of moduli masses squared (aside from the residual angular moduli and fivebrane positions to be discussed in the
next subsection) is
\eqn\modmasses{m^2_{moduli}\sim M_4^2 a g^2\sim M_4^2 {g^2\over L^6}\sim
M_4^2{1\over K^3}}
%

The lightest KK modes are those propagating in the $u,\tu$ directions, which scale like
\eqn\KKmass{m_{KK}^2\sim {1\over {\alpha'L_{u}^2}}
\sim M_4^2{1\over K^3} \left({K\over M}\right)^{1/2} }
(where we used the relation $1/\alpha'\sim M_4^2g_s^2/L^6=M_4^2 g^2$ between the string and Planck mass scales).


%
%
The
masses of the strings wound around the $L_x$ direction are of order
\eqn\windmass{m_{winding}^2\sim {L_x^2\over\alpha'}\sim
M_4^2{1\over {MK^2}}}
(This includes the effects of the $B$ field \B, which adds a term of the same order $-B G^{-1}B$ to the winding
mass squared.)

Now let us compare these scales to the Hubble scale $H$ of the de Sitter minimum:
\eqn\minU{{{\cal U}_{dS}\over M_4^4}\sim {H^2\over M_4^2} \sim \delta {b^2\over 4a}g^4 \sim \delta
{1\over K^3}}
Since $\delta$ can be tuned to be small by adjusting the precise value of $K/M$ and/or $r/M$ given $n_K/M$, this
gives a hierarchy between the de Sitter curvature scale and the scale of the moduli masses:
\eqn\Hhierarchy{H^2\ll m_{moduli}^2 ~~~~~~~ {\rm for} ~~ \delta\ll 1}
%


Of course as one adds small corrections, the precise tune in the discrete quantum numbers which one must do to
obtain small $\delta_0$ changes accordingly. As in the realistic context, it would not be possible to explicitly
tune arbitrarly finely to cancel all the loop corrections, simply because we do not know the value of these loop
corrections.

For $K\sim M$
we do not have a hierarchy between the heaviest moduli and the lightest KK modes, and between the KK modes and
the winding modes. Also, the fivebrane cores in this case are of the same order as $L_u$.  In the next
subsection, we will elaborate the model to separate these scales.

Before turning to that, in the marginal case let us discuss the question of mixing between the lightest KK modes
and the heaviest moduli.
First, note that in the nilmanifold by itself, the KK modes do not mix linearly with the moduli fields, to good
approximation. This can be seen by constructing the Laplacian on the space, but follows more intuitively from
the topology and the physics of the ``metric flux". The lightest KK modes -- i.e. those which are not
parametrically separated in mass scale from the moduli -- have no momentum in the small $x,\tx$ directions or in
the $u_1,\tu_1$ directions. Dimensionally reducing on the $x$ and $\tx$ directions, these KK modes are simply
uncharged particles on a torus with Kaluza Klein magnetic flux. Since the particles are uncharged, they have the
same spectrum as they would on a two torus, and modes of different momentum do not mix linearly.  This property
continues to hold classically after the orientifold projection is made, for the standard reason that tree-level
amplitudes for untwisted modes are inherited.  The KK5-branes do however break the translation invariance in the
$u_2,\tu_2$ directions, and their effects would need to be included in a full analysis of the moduli+KK dynamics
in the $K\sim M$ case.

\subsec{Separating the Scales}

Rather than including the KK modes in the analysis, it might be simpler to dress up the model so as to push
apart these marginally overlapping scales. Of course, additional ingredients used to do this must be introduced
in a way which does not destabilize the model. There are several approaches to this; we will suggest one method
here.

First, note that the relevant ratio of scales is
\eqn\ratio{{m_{KK}\over m_{moduli}}\sim L_x^{1/2}L^{3/2}\sim {K^{1/4}\over M^{1/4}}}
(This quantity also controls the ratio of $L_u$ to the core size of the 5-brane collection, the ratio of
lightest winding to lightest momentum masses, and the dual $10d$ string coupling.)  In our setup discussed
above, the combination of \firstmin\ and \twofirst, combined with the requirement \window, bounds the quantity
$L_x^{1/2}L^{3/2}$ to be of order $M^0$ in our parametric limit. However, as discussed above, from \newscales\
we see that if the contribution from the KK fivebrane tensions were reduced by some factor $\epsilon<1$, then
the quantity \ratio\ would be larger, of order $1/\epsilon$.  (To be clear:  we will keep $n_K=M$ for the
topological reasons discussed above; lowering the tension of the KK fivebranes would feed into the scalings
\newscales\ as if we had reduced $n_K$.)

One way to arrange this is to introduce a source of varying string coupling $e^{\phi_loc(u,\tu)}$ within the
compactification, so that the KK fivebranes (whose tensions scale like $e^{-2\phi_{loc}}$) are reduced when they
sit at a position within the compactification with increased string coupling (which minimizes their energy). Our
notation $\phi_{loc}$ here refers to the spatially varying dilaton within the compactification -- note that the
Einstein frame conversion factor \Udef\ involves the ambient $4d$ string coupling $e^\phi$, so that the effect
of an inhomogeneous dilaton on the potential contribution \Ufive\ is to rescale it by a factor
$e^{-2(\phi_{loc}-\phi)}$.

NS fivebranes provide one source of varying string coupling -- $e^{\phi_{loc}}$ increases as one moves toward
their cores.\foot{There are variants of this approach -- for D-$(p\ge 4)$-branes the coupling grows away from
the cores of the solutions, suggesting a similar mechanism with the KK fivebranes drawn to positions in between
added sets of D-branes.} Moreover, in our setup, NS fivebranes which are wrapped on the $x,\tx$ directions each
introduce parametrically less potential energy than the leading terms \Ufull:
\eqn\Ufivenew{{\cal U}_{NS5}\sim g^2{n_5 L_x^2\over L^6}}
Specifically, comparing this to the curvature term in \Ufull\ using \twonext\firstmin\ we see that as long as
$n_5\le M$, adding a set of such NS5-branes provides a term in the potential which is at or below the scale of
the above ingredients. At the same time, it provides a varying string coupling within the compactification,
which we may be able to use to reduce the KK5-brane contribution to the potential.

Since the KK 5-branes wrap one of the $u,\tu$ directions, the configuration of interest is one in which they
skirt the cores of one or more clouds of NS5-branes as they stretch across them, lowering their energy by
passing through regions with lower string coupling. Each NS 5-brane is localized in the $u,\tu$ directions, but
they may be distributed so as to minimize the energy of the whole configuration.  We will now estimate whether
this effect can be significant in our background.

The varying string coupling in the NS5-brane solution is
\eqn\NSgs{e^{2\phi_{loc}(r)}=g_s^2+{\sum_{i=1}^{n_5}{\alpha'\over{2\pi^2 (r-r_i)^2}}}}
where $r$ is the radial coordinate, the transverse string frame metric being
\eqn\NSmet{e^{2\phi_{loc}(r)}(dr^2+r^2d\Omega^2).}
To check if the NS5 branes can significantly change the string coupling -- and hence KK5 tension -- let us start
from the previous results and estimate the magnitude of the effect in that configuration.  From \NSgs\ and the
result \gstring\ that without the present effect, $g_s\sim 1/K^{1/2}\sim 1/M^{1/2}$, we see that within a radial
position $r_*$ of order $n_5^{1/2}M^{1/2}$, the string coupling is affected significantly by the NS5-branes.
Moreover, the metric \NSmet\ at most increases the minimal length of the path traced by the fivebrane in the
$u,\tu$ directions by one power of $e^{\phi_{loc}}$, which cannot cancel the effect of the tension (which is
quadratic, of order $e^{-2\phi_{loc}}$).  Also, the NS5-brane solution does not warp the string frame metric in
the directions along its worldvolume, including $x,\tx$. Since the KK5-brane core and $L_u$ are of size
$M^{1/2}$ in the original solution, we see that introducing any NS5-branes would begin to reduce the tension of
the whole KK5-brane set.

Since the KK5-branes are extended in one direction in $u,\tu$, it is perhaps more natural to distribute our
$n_5$ NS5-branes along this line, putting them out away from the origin of their moduli space. (This is not
crucial to get an effect from them, as we just saw, but it is the most symmetric configuration and one to which
the system could settle as it minimizes its energy locally.) Since the relevant scales only overlapped
marginally in the above solution, any reduction of the tension of the KK5-branes is sufficient to push the setup
in to a regime where the moduli masses are lighter than all KK masses and where the other related scales
discussed above are also separated.  Obviously this collection of KK 5-branes and NS 5-branes is rather
complicated to analyze in detail, but because they carry different charges we do not expect any catastrophic
annihilation mode.

As in the above discussion of KK fivebranes, there are possible topological consistency conditions which may
constrain $n_5$ to be a multiple of $M$, depending on the application of \VZ\ to the $m_0 B_{x\tx}$ contribution
to the generalized fluxes \Ffull.  Because of this, we have checked that the case $n_5=M$ remains consistent
with the window \window; it contributes a new pair of terms going like $L_x^2$ and $1/L_x^2$ which can be
analyzed in the same way as we did \twonext\ above.  The case $n_5=M$ is also consistent with our geometry --
even if all NS5-branes were together, the coupling \NSgs\ does not grow to order 1 until $r\sim n_5^{1/2}$, and
hence before taking into account the improvement in the scalings from the varying dilaton, their core would be
about the size of the transverse space (as was true for the original KK fivebranes).

\subsec{Angular and Fivebrane Moduli}

Having addressed the runaway moduli from the volumes and dilaton, let us return to the angular moduli.  These
fall into two categories:

\item{(1)}  Angular moduli sourced by the ingredients listed above
\item{(2)}  Angular moduli not sourced by the ingredients listed above.

\noindent Those in category (1) must be analyzed on the same footing as the moduli $L, L_x, g$ listed above, to
ensure that they do not turn on to large enough values to potentially destabilize the dS minimum.  Those in
category (2), such as some of RR axions we will discuss, will not destabilize the dS minimum wherever they end
up, and will generically be lifted by higher order, lower-scale corrections.  Many of the moduli in both
categories could be projected out by an orbifold version of the above construction (with appropriate corrections
to the order one factors in the volume and potential). However, it is also interesting to pursue their
stabilization without using that crutch, since many of the existing ingredients provide the requisite forces in
a natural way.

First, recall that the continuous Wilson line moduli $b_{I\tilde J}$ coming from the $B$ field \B\ are lifted by
the $F_2^2$ term, since $F_2=m_0B+dC_1$.  The metric flux renders the components $B_{u\tx}=-B_{\tu x}$ discrete
because the $x,\tx$ circles are torsion 1-cycles.  (These are some of the discrete Wilson lines discussed in
\S3.2.)


Next, consider the metric moduli.  As discussed around \metmod, the curvature alone gives positive mass squareds
to ${\cal G}_{x\tx},{\cal G}_{u_i\tu_j}$, and ${\cal G}_{\tx\tu_j}={\cal G}_{x u_j}$.  However, we must consider
other contributions to their potential from the other ingredients in the construction.


Let us discuss first the angular degree of freedom $\eta$ defined in \defeta\gammaeta.  Since this comes from
$G_{x\tx}$, this particular modulus is not easy to project out by an orbifold without also projecting out the
$B_{x\tx}$ contribution which played an important role above. Several of the ingredients we have specified
depend on this angle, as explained previously, in the discussion following \defeta\gammaeta. Including these
effects, we obtain a potential energy of the form
\eqn\Ufulleta{\eqalign{& { {\cal U}\over M_4^4}  =ag^2-bg^3+cg^4\cr &= g^2\left([1+(\eta-\eta_{q'})^2+\dots]M^2
{L_x^4\over {2 L^6}}+(4 {n_{K}\over\sqrt{\eta}}) {L_x^{5/2}\over L^{9/2}}+{3 p^2\over {2\eta (\alpha')^2
L^6}}\right)-g^3\left(2{\sqrt{2\over {\eta \alpha'}}}|p m_0|\right)\cr & +g^4\alpha'\left({m_0^2\over
4}L^6+4\pi^4m_0^2\left({q\over M}\right)^2{L^6\over L_x^4}
+\left({r\over M}\right)^2{16\pi^4m_0^2L^3\over{\eta L_x}}+\left({r\over M}\right)^4{2^8\pi^8 m_0^2\over{
L_x^2}} +{K^2\over{4 L^6(\alpha')^6}}\right)\cr }}
Correspondingly, the quantity \combo\ becomes
\eqn\comboeta{\eqalign{& {4ac\over b^2} \equiv 1+\delta(L,L_x,\eta) \cr & ={\eta(\alpha')^2\over{2
p^2m_0^2}}\left([1+(\eta-\eta_{q'})^2+\dots]M^2 {L_x^4\over {2 L^6}}+(4 {n_K \over \sqrt{\eta}}) {L_x^{5/2}\over
L^{9/2}}+{3 p^2\over {2\eta (\alpha')^2 L^6}}\right)\cr &\times\left({m_0^2\over 4}L^6+4\pi^4m_0^2\left({q\over
M}\right)^2{L^6\over L_x^4}
+\left({r\over M}\right)^2{16\pi^4 m_0^2 L^3\over{\eta L_x}}+\left({r\over M}\right)^4{2^8\pi^8 m_0^2\over{
L_x^2}} +{K^2\over{4 L^6(\alpha')^6}}\right)\cr}}
As it stands, in \comboeta\ there is a runaway direction in which $\eta\to 0$ with $L^3L_x\propto 1/\eta$. (In
this limit $\eta\to 0$, we find also that the curvature potential has a term proportional to $1/\eta$.)  But for
sufficiently large $\eta_{q'}$, of order 1, the stabilizing mass from the curvature term competes with the
tadpoles from the other ingredients; we expect the two to balance to yield a {\it local} minimum in this
direction.  There is certainly a local minimum in the $\eta$ as well as in the $g,L,L_x$ directions, at the
point where $\eta\sim 1\sim\eta_q$ and $g,L,L_x$ sit at the minimum derived in \S3.4.  The remaining question is
whether the other moduli can mix with $\eta$ significantly enough to introduce large enough off-diagonal
contributions to the mass matrix to destabilize the system.  Generically, near $\eta=1$ the second derivatives
$L^2\del_L^2,L_x^2\del_{L_x}^2,g^2\del_g^2,\del_\eta^2,$ and the mixed terms $L\del_L\del_\eta,\dots$  acting on
$(4ac/b^2)$ are of the same order (that of the typical term in $4ac/b^2$).  In a similar way to the way we tune
the cosmological constant by picking $K/M$, we can pick $r/M$ and if necessary the coefficients of additional
fluxes we did not use yet (such as other two-form or four-form fluxes) to help tune the off-diagonal elements of
the mass matrix to be smaller than the diagonal elements if necessary.  The degree to which this extra tuning is
required depends on the order one coefficients involved in determining these derivatives, which we have not
explicitly calculated here.


Let us now discuss the other angular moduli, which have some similar features.  In general, turning on angles in
one direction $G_{\mu\nu}$ at fixed volume increases the size of the cycle in either the $\mu$ or $\nu$
direction. For example, turning on ${\cal G}_{u_1\tu_1}$ at fixed volume of the two-torus in the $u_1,\tu_1$
directions increases the size of the cycle wrapped by the O6-plane and the $KK5$ branes, and it decreases the
size of the 3-cycle threaded by the $H$ flux. Let $\rho_{1\tilde{1}} \sqrt{2} L_{u_1}$ denote the size of the
cycle in the $u_1+\tu_1$ direction (normalized so that $\rho_{1\tilde{1}}=1$ for the diagonal metric \symmmet).
The O6-plane energy and the KK5-brane energy each get multiplied by $\rho_{1\tilde{1}}$, and the $H$ flux term
gets a factor of $\rho_{1\tilde{1}}^2$.

Taking these effects into account for all angles (defining $\rho_{\mu\nu}$ similarly to how we just defined
$\rho_{1\tilde{1}}$), the quantity $4ac/b^2$ \combo\ takes the form
\eqn\comborho{\eqalign{& {4ac\over b^2} \equiv 1+\delta(L,L_x,\rho) \cr & ={(\alpha')^2\over{
\theta_O^2(\rho_{\mu\nu}) (2p^2m_0^2)}}\left(\theta_{\cal R}(\rho)M^2 {L_x^4\over {2
L^6}}+\theta_{KK}(\rho_{\mu\nu})(4 n_K) {L_x^{5/2}\over L^{9/2}} +{\theta_H(\rho_{\mu\nu}) 3 p^2\over
{2(\alpha')^2 L^6}} \right)\cr &\times\biggl({m_0^2\over 4}L^6+4\pi^4m_0^2\left({q\over M}\right)^2{L^6\over
L_x^4}
+ \theta_{r,2}(\rho)8\pi^4 m_0^2\left({r\over M}\right)^2{L^3\over{L_x}}+\theta_{r,4}(\rho)\left({r\over
M}\right)^4{2^6\pi^8 m_0^2\over{ L_x^2}}+{K^2\over{4 L^6(\alpha')^6}}\biggr)\cr}}
%
where we have schematically indicated by functions $\theta(\rho)$ the dependence on $\rho_{\mu\nu}$ of those
ingredients which are sensitive to these angles.

There are two types of angles in our problem: those which are sourced by the O6-plane and H flux, analogously to
the $G_{x\tx}$ mode discussed above (${\cal G}_{u_1\tu_1}$ and ${\cal G}_{u_2\tu_2}$), and those which are not
(${\cal G}_{u_1u_2}={\cal G}_{\tu_1\tu_2}$, ${\cal G}_{u_1\tu_2}={\cal G}_{\tu_1 u_2}$, and ${\cal
G}_{x,\tu_j}={\cal G}_{\tx,u_j}$).

Let us start with the former case. As in our discussion of $G_{x\tx}$, the mass squared introduced by the
curvature has the right shape and order of magnitude to provide a local minimum when balanced against the
tadpoles from the O6 and $H$ flux, depending on order one coefficients and on discrete parameters that can be
tuned.  We have not analyzed these coefficients in detail, so let us mention two other methods for stabilizing
$\rho_{u_1\tu_1},\rho_{u_2,\tu_2}$.  First,
we can consider adding an additional sector of KK fivebranes to avoid the $\rho\to\infty$ direction. Namely, a
KK5-brane wrapped on the $x,\tx$ direction, and with fiber circle $u_1+\tu_1+u_2+\tu_2$ gives a potential term
scaling like $g^2 \rho^2 L_x/L^3$.  This prevents the potential runaway limit of \comborho\ to large $\rho$, and
the new ingredient is at most of the same order as the previous contributions to the potential, and so can be
added consistently.
%
Alternatively, it is possible to enforce $\rho=1$ by an orbifold. One example is a $\IZ_2$ orbifold under which
$u_j\to -u_j,\tu_j\to -\tu_j,x\to x, \tx\to \tx$ (either combined with a transverse shift to remove twisted
moduli, or with extra fivebranes wrapped around the blowup cycles, carrying worldvolume gauge flux to stabilize
their sizes). This introduces an O6 fixed plane in the directions $u_j-\tu_j, x+\tx$, whose linear potential for
${\cal G}_{u_1\tu_1}$ and ${\cal G}_{u_2\tu_2}$ cancels against that of the original O6-plane in the symmetric
configuration \symmmet, making it manifest that the curvature mass term \SS\ suffices to lift this angular
direction. (Various other factors in the tadpole cancellation condition and in the potential also change in the
orbifold case, and the contributions proportional to $r/M$ -- which were not crucial in the stabilization above
-- are projected out.  A preliminary check of the coefficients resulted in parameters still consistent with the
window \window; it would be useful to systematically study different orbifold space groups with regard to the
question of the constant terms in \combo\ resulting from the normalized volume and tadpole cancelation
conditions.)

The angular moduli which are not sourced by the O6-plane and $H$ flux are also stabilized by curvature in the
case ${\cal G}_{u_1\tu_2}={\cal G}_{\tu_1 u_2}$.   In the case ${\cal G}_{u_1u_2}={\cal G}_{\tu_1\tu_2}$ they
are stabilized by the KK fivebranes: turning on angles between the $u_1,\tu_1$ and $u_2,\tu_2$ directions at
fixed volume increases the lengths of the cycles the KK5-branes are wrapped on, while not affecting the volume
of the cycle wrapped by the O6-plane or the cycle threaded by the $H$ flux. (A similar mechanism arises in
intersecting brane models \IBM.).




This leaves us with ${\cal G}_{u_j\tx}={\cal G}_{x\tu_j}$ (if the $\IZ_2$ orbifold option described above is not
taken -- this would simply project them out). These are metric Wilson lines. Recall that the metric flux lifts
the $B_{u\tx}=-B_{\tu x}$ modes. $H$ flux similarly lifts the modes ${\cal G}_{u_j\tx}={\cal G}_{x\tu_j}$, by
$T$-duality.  In our case, the $H$ flux potential is minimized at one of two values, since we took the $H$ flux
quantum number to be 2. That is, these metric Wilson lines are discrete. These modes are thus similar to those
discussed above, but with the positive mass squared contribution coming from $H$ flux rather than from the
curvature term.

The RR axions are either fixed by virtue of the Chern-Simons terms in \Ffull, or contribute subdominantly to the
potential (hence falling in category (2) above). There are three components of $C_3$ invariant under the
orientifold action \Oplane. They are dual to the components of $H$ indicated in \Ocycle, and are stabilized by
the $H$ flux, since
\eqn\FsixC{\tilde F_6=dC_5-C_3\wedge H_3+{m_0\over 6}B\wedge B\wedge B}
%
Similar comments apply to $C_1$ and $C_5$:  components
which are not projected out by the orientifold action or lifted by the $|\tilde F_4|^2$ terms are fixed by
higher order effects which generate the axion potential.


%
%

Finally, let us return to the motion moduli of the KK fivebranes.  As discussed above, the discrete torsion
requires the $M$ branes to sit together, projecting out their relative motions. Given the mechanism suggested in
\S3.6, the size of the set of $n_K=M$ KK fivebranes is parametrically smaller than $L_{u}$, and in all versions
of the construction it is no larger than the transverse space. The overall position of the KK5-branes is
inconsequential to the stabilization of the runaway moduli, and hence directions which are not fixed by the
curvature potential are in category (2) above; again, if we invoke the method of \S3.6, then the position of the
KK 5-branes is localized near the source of enhanced $e^{\phi_{loc}}$. There are no isometries in the transverse
directions to the KK fivebranes, so in any case their position moduli will be lifted by effects to do with the
ambient curvature of the space transverse to the KK5-branes.

\subsec{Numerical solution}

We showed analytically above that a local minimum of \Ufull\ exists for appropriate choices of discrete quantum
numbers. We have checked this numerically using mathematica.  In doing so it was again useful to follow the
procedure used above, first finding a minimum of $4ac/b^2$ at some $L=L_0,L_x=L_{x0}$ (tuning the discrete
quantum numbers $f_6,M$ to arrange for the minimum value $\delta_0$ of $1-4ac/b^2$ to be close to 0).  Next, we
minimized ${\cal U}(g,L_0,L_{x0})$ with respect to $g$. Then searching near that point for a minimum in all
directions yields the expected solution. As a specific example, with $n_K=M=10$, $f_6=80$, $q=1$, and $r=1$ the
minimum of $4ac/b^2$ is at approximately $1.0003$
(so $\delta_0=0003\ll 1$ is very small, putting us in the regime of applicability of the analysis in \S2). The
potential is minimized at ${\cal U}/M_4^2\approx 10^{-13}$ with
$g\approx 0.00015,L \approx 15.3, L_x \approx 2.1$.

Note that in this solution, with $M$ taken to be 10, $L_x$ is {\it not} substringy in size, the string coupling
$g_s$ is of order 1/2, and the T-dual string coupling is also not strong. Scaling $M$ up pushes $L_x$ down as
discussed in the text, but at modest finite values of the parameters such as those given here one can obtain
less extreme behavior.

Plotting the potential ${\cal U}/M_4^4$ in the $g$ direction yields

\fig\gstab{The potential in the $g$ direction.} {\epsfxsize2.0in\epsfbox{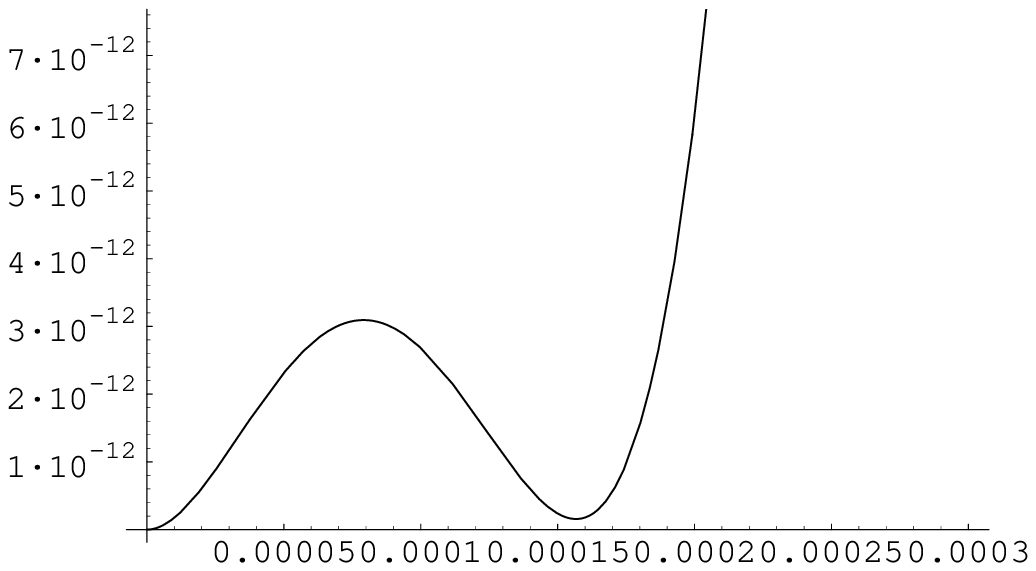}}
%

and in the $L$ and $L_x$ directions we obtain (note that the horizontal axis is not placed at zero)

\fig\Lstab{The potential in the $L,Lx$ directions.} {\epsfxsize2.0in\epsfbox{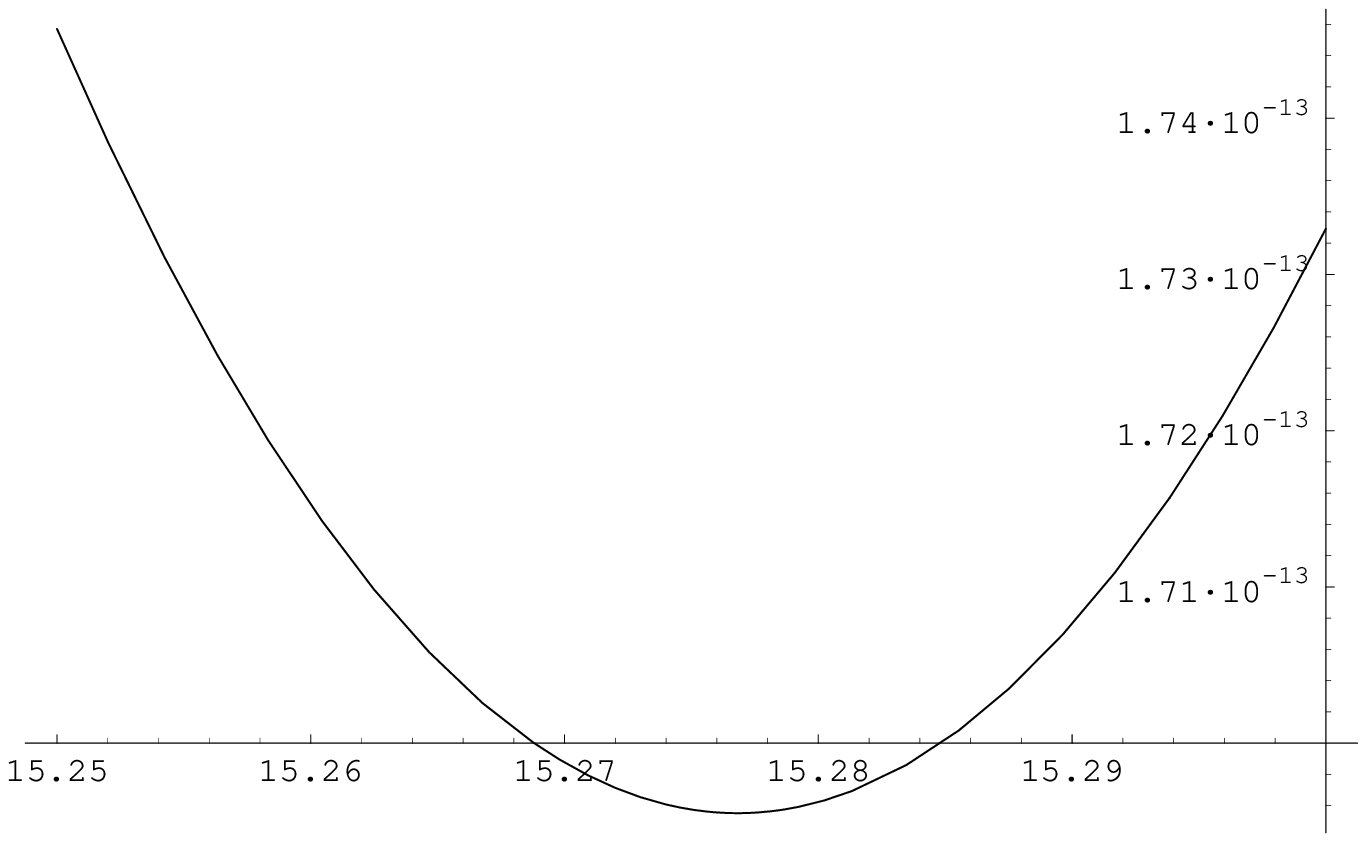}}
{\epsfxsize2.0in\epsfbox{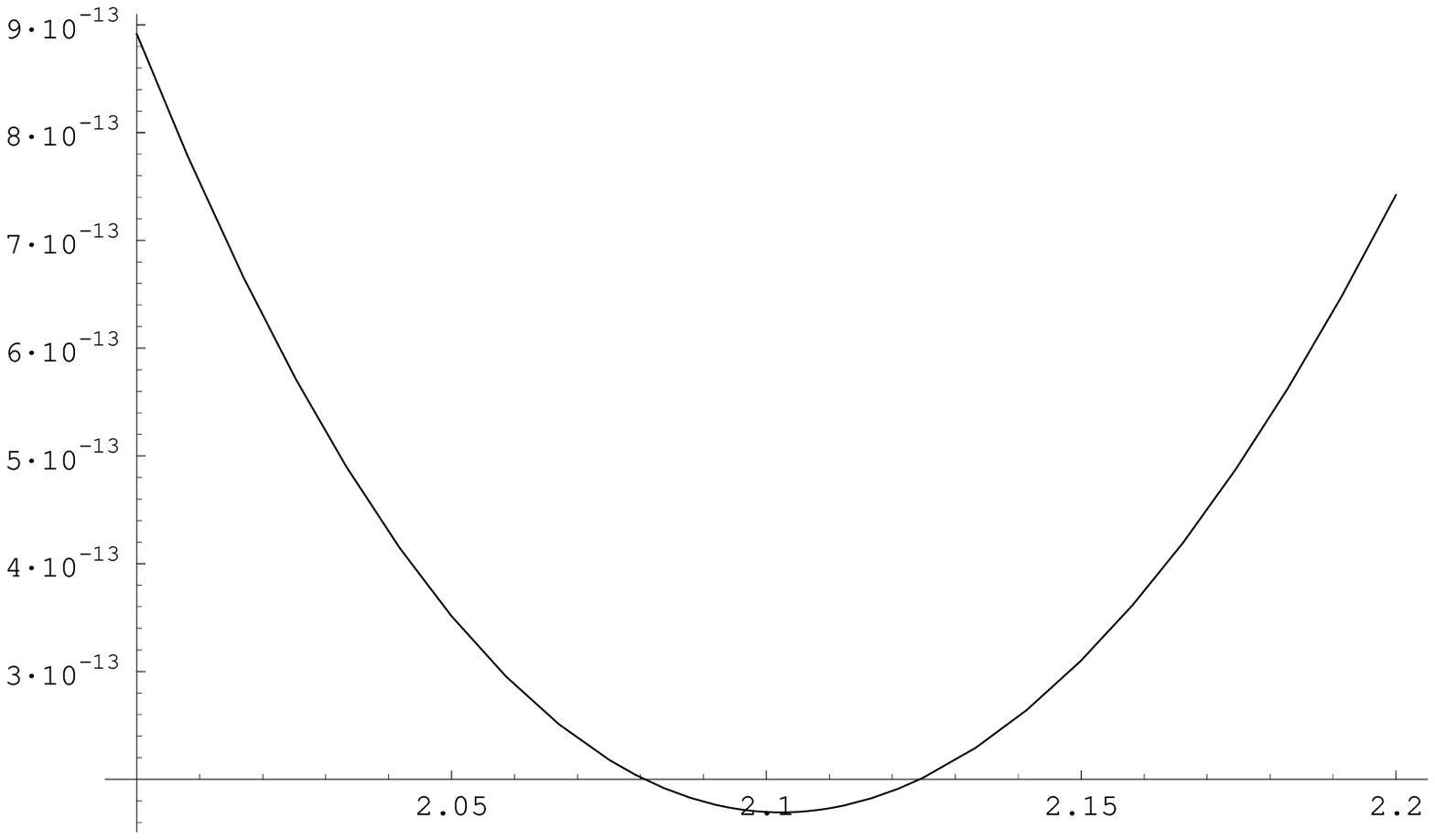}}

Note that this numerical analysis does not explicitly include the angular variables discussed in \S3.7.

\subsec{Metastability}

As emphasized in \refs{\SilversteinXN,\KachruAW}, models of de Sitter which are obtained in a weak coupling
regime are only metastable.\foot{For a recent comparitive study of decays in a subset of string theoretic dS
models, see for example \WestphalXD.} In the present case, the decay yields a ten-dimensional generalization of
a Bianchi cosmology, with different directions evolving anisotropically -- the $x,\tx$ directions shrinking and
the others expanding. As discussed above, a nilmanifold in vacuum can be T-dualized to an expanding space with
$H$ flux \KachruSK, but since the $B_{x\tx}$ field is nontrivial in our solution the element of the T-duality
group which is relevant at a given radius is not the simple one considered in \KachruSK.  The question of
whether one can or cannot T-dualize to large radius, perhaps via a time-dependent T-duality cascade\foot{I thank
S. Kachru for this suggestion. These questions are also related to those analyzed in \dualsing.}, is an
interesting one.

\newsec{Toward Simple de Sitter Sol-utions}

We expect similar solutions from a compactification on a product of two Sol 3-manifolds; let us sketch the
analogues of the steps given above for the nilmanifold case.  The sol 3-geometry is
\eqn\Solmet{\eqalign{ds^2 & = \alpha'\left(L_y^2(e^{2z}dy_1^2+e^{-2z}dy_2^2)+L_z^2dz^2\right)\cr & =
\alpha'\left(L_y^2\omega_1^2+L_y^2\omega_2^2+L_z^2\omega_3^2\right)}}
where $\omega_1=L_ye^zdy_1,\omega_2=L_ye^{-z}dy_2$, and $\omega_3=L_zdz$.  This geometry has three independent
isometries consisting of shifts of $y_1,y_2$ and shifts of $z$ combined with rescalings of $y_1,y_2$.

Compact solmanifolds ${\cal S}_3$ are obtained by projecting \Solmet\ by a discrete subgroup of the isometry
group. As one moves around the $z$ direction, the $\tau$ parameter of the $T^2$ in the $y_1,y_2$ directions
undergoes an $SL(2,\IZ)$ transformation. This is analogous to \compactification\ except that here the
$SL(2,\IZ)$ transformation must be more general than $\tau\to\tau+1$ in order to provide a consistent
compactification; it shrinks the torus exponentially in the $y_1$ direction and expands it exponentially in the
$y_2$ direction. (Related to this, the solmanifold has a rich fundamental group, which is of exponential
growth.)

As with the nilmanifold, the sol manifold has fewer massless moduli than the corresponding $T^3$: the relations
\eqn\holrelns{d\omega_1=\omega_3\wedge\omega_1 ~~~~ d\omega_2=-\omega_3\wedge\omega_2}
mean that the homology groups $H^1$ and $H^2$ are each reduced by two dimensions as compared to a torus.

The scalar curvature of ${\cal S}_3$ is $-2/L_z^2=-2L_y^4/L^6$ where $L^6=L_y^4L_z^2$.  Upon compactification to
four dimensions on a product of two solmanifolds this will lead to a positive curvature potential analogous to
\potR, with discrete parameters analogous to $M$ in \potR\ which have to do with the choice of $SL(2,\IZ)$
element used in the compactification procedure. As in the case of the nilmanifold, the rich topology of ${\cal
S}_3$ provides a place for branes to wrap and a potential source of fractional Chern-Simons invariants.  All
this again leads to a four dimensional scalar potential analogous to \Ufull, with the role of $x$ played by
$y_1,y_2$ and the role of $u_1,u_1$ played by $z$. It would be interesting to flesh this out explicitly to see
if again the discrete parameters available are sufficient to tune the system into the range \dScond.

\newsec{Discussion}

In this work, we proposed a relatively simple and explicit class of de Sitter models in string theory. We showed
how a few ingredients suffice to produce a potential for moduli which exhibits metastable minima at positive
vacuum energy, seven independent terms being involved in the basic stabilization of $g, L$, and $L_x$.
Clearly an important direction for further work is fleshing out further the methods in \S3.6 and \S3.7\ for
stabilizing the angular moduli and for separating the scales. A convenient feature of the background is its weak
curvature and 10d string coupling, and its $10d$ supersymmetry, which make possible a controlled analysis of the
moduli and the KK and winding modes. On the other hand, the SUSY breaking effects of the curvature and KK
fivebrane configuration facilitate moduli stabilization by introducing useful competing forces (which would
vanish in the lower-energy SUSY models based on Calabi-Yau manifolds with the subset of ingredients analyzed in
the no-go theorem of \mit).

One of the main general lessons is that ``metric flux" and wrapped KK 5-branes yield forces whose dependence on
the moduli is appropriate to ``uplift" the potential for the runaway moduli $g,L,L_x$ in the type IIA $AdS_4$
solutions of the sort studied in \DeWolfeUU.  The most complicated aspect of the specific models is probably the
5-brane dynamics.

One natural question is whether a version of this mechanism exists in which lower energy supersymmetry is
preserved.  In \VZ\ it was suggested that various ``metric fluxes" could cancel the charges of $KK$ fivebranes
as well as $NS5$-branes.  Such a construction could be analogous to the way fluxes cancel the $O3$-plane charges
in the type IIB models of \refs{\GiddingsYU,\KachruAW}.  Combining this idea of \VZ\ with the mechanism
described here might be a concrete place to seek models with lower scale supersymmetry breaking.

It would be interesting to apply our construction to the problem of explicitly modeling inflation in string
theory (for recent reviews see \Inflationreviews).  One question is whether our fractional Chern-Simons
invariants could also help tune the inflationary $\epsilon$ and $\eta$ parameters to be small.  It might also be
interesting to introduce particle physics sectors to these models, perhaps using stretched D-branes within the
bulwark of KK5-branes to form brane constructions of the relevant field theories.

Explicit models of de Sitter (and also anti de Sitter, if we reduce $K/M$) may facilitate the derivation of
concrete holographic duals.  Some progress toward a general framework for duals of metastable de Sitter space
have appear in \dualgen.  Ideas for unveiling the degrees of freedom of the dual out on its approximate moduli
space can be found in \junctions, by trading the flux for branes in analogy with the construction of the Coulomb
branch in more familiar versions of the AdS/CFT correspondence. In the present case, one might trade the $H$
field and metric flux for $NS5$-branes and KK monopoles.

\centerline{\bf{Acknowledgements}} It is a pleasure to thank B. Florea, T. Grimm, and A. Tomasiello for very
useful discussions; O. Aharony, D. Green, S. Kachru, L. McAllister, W. Taylor and A. Westphal for useful
discussions as well as very useful comments on a draft; and A. Lawrence, A. Maloney and D. Starr for ongoing
discussions on the ``fundamental" physics of compact target spaces and their T duals. This research is supported
in part by NSF grant PHY-0244728, and in part by the DOE under contract DE-AC03-76SF00515.



\listrefs

{\epsfxsize5.0in\epsfbox{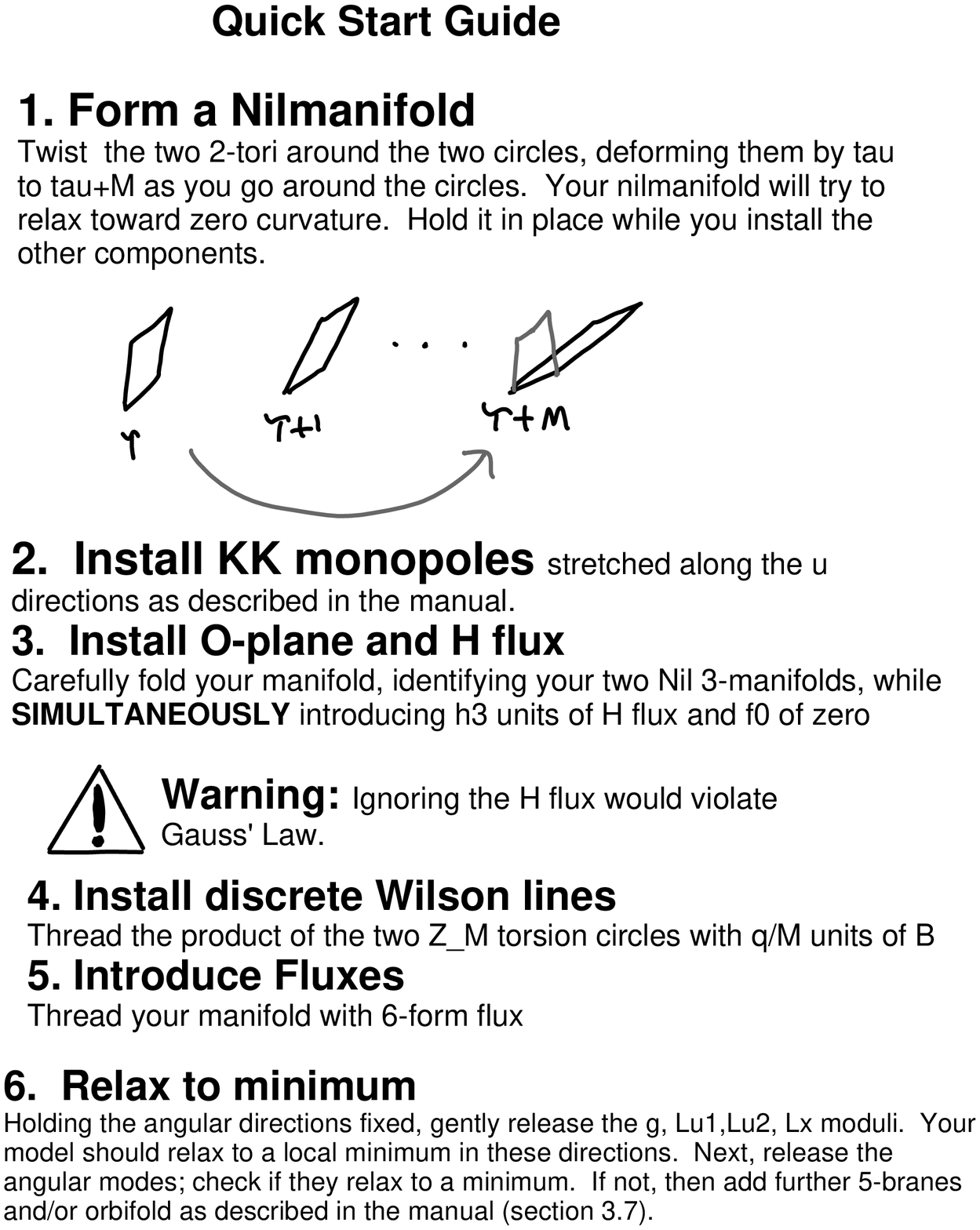}}

\end